\documentclass[aps,preprint,nofootinbib]{revtex4-2}

\usepackage[top=2.5 cm, bottom=2.5 cm, left=2.5 cm, right=2.5 cm]{geometry}

\usepackage[utf8]{inputenc}
\usepackage{amsfonts}
\usepackage{color}

\usepackage{amssymb}
\usepackage{amsmath}
\usepackage{stmaryrd}
\usepackage{latexsym}
\usepackage{tcolorbox}
\usepackage{multirow}

\usepackage{xcolor}
\definecolor{myurlcolor}{HTML}{08457E}
\definecolor{mylinkcolor}{HTML}{2A52BE}
\definecolor{mycitecolor}{HTML}{E30022}

\usepackage{float}
\usepackage[colorlinks, linkcolor=mylinkcolor, citecolor=mycitecolor, urlcolor=myurlcolor, linktocpage=true]{hyperref}

\def\equationautorefname~#1\null{(#1)\null}
\def\tableautorefname~#1\null{(#1)\null}
\def\figureautorefname~#1\null{(#1)\null}
\def\sectionautorefname~#1\null{(#1)\null}

\let\origref\autoref
\def\autoref#1{\textbf{\origref{#1}}}

\let\origcite\cite
\def\cite#1{\textbf{\origcite{#1}}}

\usepackage{tikz}
\usepackage{pgfplots}
\pgfplotsset{compat=1.15}

\usepackage{titlesec}
\titleformat*{\section}{\centering\small\bfseries\scshape}
\titleformat*{\subsection}{\small\bfseries\scshape}
\titleformat*{\subsubsection}{\small\bfseries\scshape}

\usepackage{array}
\newcommand{\PreserveBackslash}[1]{\let\temp=\\#1\let\\=\temp}
\newcolumntype{C}[1]{>{\PreserveBackslash\centering}p{#1}}
\newcolumntype{R}[1]{>{\PreserveBackslash\raggedleft}p{#1}}
\newcolumntype{L}[1]{>{\PreserveBackslash\raggedright}p{#1}}

\def\S{{\cal S}}  
\def\R{{\cal R}}     
\def\V{{\cal V}}     
\def\T{{\cal T}}     
\def\P{{\cal P}}     
\renewcommand{\d}{\mathrm{d}}
\newcommand{\Mpl}{M_{\mathrm{Pl}}}
\newcommand{\mpl}{m_{\mathrm{Pl}}}

\newcommand{\sfrac}[2]{\dfrac{\,#1\,}{\,#2\,}}

\newcommand{\der}[2]{\sfrac{\d #1}{\d #2}}
\newcommand{\derp}[2]{\sfrac{\partial #1}{\partial #2}}

\newcommand{\mf}[1]{\mathbf{#1}}
\newcommand{\nb}{\nabla}
\newcommand{\sg}{\sigma}
\newcommand{\dt}{\delta}

\newcommand{\lb}{\lambda}
\newcommand{\al}{\alpha}
\newcommand{\bt}{\beta}
\newcommand{\p}{\partial}

\let\oldsqrt\sqrt
\def\sqrt{\mathpalette\DHLhksqrt}
\def\DHLhksqrt#1#2{%
	\setbox0=\hbox{$#1\oldsqrt{#2\,}$}\dimen0=\ht0
	\advance\dimen0-0.4\ht0
	\setbox2=\hbox{\vrule height\ht0 depth -\dimen0}%
	{\box0\lower0.4pt\box2}}

\begin{document}

\title{The Effect of Non-minimally Coupled Scalar Field on Gravitational Waves from First-order Vacuum Phase Transitions \vspace{10mm}}

\author{A.\ Savaş Arapoğlu}
\email{arapoglu@itu.edu.tr}

\author{A.\ Emrah Yükselci}
\email{yukselcia@itu.edu.tr}

\affiliation{\vspace{5mm}Istanbul Technical University, Faculty of Science and Letters, Physics Engineering Department, 34469, Maslak, Istanbul, Turkey \vspace{2cm}}

\begin{abstract}
We investigate first-order vacuum phase transitions in the presence of a non-minimally coupled scalar field starting with the coupling effect on the initial dynamics of phase transitions by defining an effective potential for the scalar field and then performing three dimensional numerical simulations to observe any possible distinction in gravitational wave power spectrum. Although we give a description of the model with the expanding background, in this particular paper, we exclude the scale factor contribution since we primarily focus on the immediate impact of the non-minimal coupling at initial phases of the transition. Even in this case we found that this modification has discernible effect on the power spectrum of the gravitational wave energy density.
\end{abstract}

\maketitle
\newpage
\raggedbottom

\section{INTRODUCTION\label{sec:intro}}

Although it was originated from an astrophysical event, merging of two black holes, the first direct detection of gravitational waves (GWs) \cite{First_GW_detection} has a special importance of examining the early Universe as well. The processes in the early Universe have some imprints on the ``Cosmic Microwave Background Radiation” \cite{CMB,PlanckCMB} which can be considered as the picture of the Universe when its age is approximately 380,000 years. But, interactions at the end of such processes may lead these imprints become weak. However, weakly interacting GWs preserve the information related to their production. Therefore, GWs produced as a result of the early Universe processes have the opportunity of encoding direct information about such an era of the Universe with their possible detection through the future-planned space-based GW detectors \cite{LISA,LISA_rev,DECIGO,TAIJI,TIANQIN,BBO} that are supposed to have the precision to a certain extent for such observations. 

First-order cosmological phase transitions (PTs) are one of those phenomena that could possibly take place in the early Universe through the bubble nucleation mechanism and could be source to GWs carrying information about the event itself. In spite of the fact that the standard model of particle physics predicts the transitions in the early Universe, such as the electroweak PT and the quark-hadron PT, as crossovers, but not a first-order PTs \cite{Kajantie1996_1,Kajantie1996_2}, such an incident could be seen in many extensions of it (See e.g. \cite{Mazumdar2019,Hindmarsh2020,Weir2017} and references therein). Therefore, it is noteworthy to analyze the outcomes of well-motivated models.

Theory of the vacuum PTs through the bubble nucleation in flat space-time were studied in Refs.\ \cite{Coleman1977_1, Coleman1977_2} and \cite{Linde1980, Linde1983} at zero and finite temperatures, respectively. Gravitational effects, on the other hand, were investigated in Ref.\ \cite{Coleman1980}. In flat space-time it was found that the most probable bubble profile is $O(4)$ symmetric configuration \cite{Coleman1978}. However, to the best of our knowledge, there is no proof of that concept for a curved space-time although in practice it is common to assume the same structure with the one in flat space-time.

The gravitational waves originated from a first-order PT are sourced by the shear stress forming after the collision of bubbles losing their symmetrical structures. In order to investigate the gravitational radiation in such an incident ``envelope approximation" was developed by neglecting the overlapping regions between the colliding bubbles \cite{Turner1992_env}. Recently, this approximation has been tested by three dimensional numerical simulations in Refs.\ \cite{Cutting2018,Cutting2019,Cutting2020} and it has been shown that the overlapping regions of collided bubbles are important in a sense that they give rise to oscillations in the scalar field distribution which are observed as a frequency peak proportional to its mass in GW power spectrum. Hence, it is crucial to run computer simulations in order to obtain more accurate results especially when investigating the long-time effects. On the other hand, there are analytical approximations which are capable of predicting the power spectra of GW energy density from bubble collision phase \cite{Caprini2007,Jinno2017,Zhong2021}.

In curved space-time the quantum corrections to the scalar field require the inclusion of a non-minimal coupling term in the action \cite{chernikov1968,callan-etal1970,birrell-davies1980,birrell-davies-1982} and with this respect that model falls under the well-motivated theories that we have mentioned before. Bubble nucleation mechanism through the non-minimally coupled scalar field were studied in different contexts before \cite{Lee2006,Salvio2016,Czerwinska2016,Rajantie2016}. In this work, we will study the effect of the non-minimal coupling on GW power spectrum by performing various three dimensional simulations and compare the results with the previous works in the literature. For this particular paper, although we will initially give a whole description of the model in an expanding Universe, we will neglect the impact of the scale factor on the equations of motion when performing the numerical simulations, since we will be interested in the short-time effects for now.

The plan of the paper is as follows: In Sec.\ \autoref{sec:main_equations} we give the main equations which will form the base for the numerical studies. The effect of the non-minimal coupling terms on the shape of the potential is investigated in Sec.\ \autoref{sec:effective_potential}. Then, the initial bubble profile, i.e. the bounce solution, is examined in Sec.\ \autoref{sec:bubbleProfile} by comparing the results with the previous works. The model for simulations is described in Sec.\ \autoref{sec:modelForSimulations} followed by Sec.\ \autoref{sec:numerical_methods} containing the numerical methods. Then, the results are given in Sec.\ \autoref{sec:results} and final remarks are discussed in Sec.\ \autoref{sec:conclusion}.

\section{MAIN EQUATIONS\label{sec:main_equations}}

In this section we give the equations that will be used throughout this work. We begin with the action for the non-minimally coupled scalar field as follows
\begin{equation}
	\S = \int d^4 x \sqrt{-g} \, \bigg[ \sfrac{1}{2} \big( \Mpl^2 + \xi \phi^2 \big) \R - \sfrac{1}{2} \, \nb^\sg \phi \, \nb_{\!\!\sg} \, \phi - V(\phi) \bigg]
\label{eq:sttaction}
\end{equation}
where $\xi$ is the coupling constant, $V(\phi)$ is self-interacting potential, and $\Mpl^{-2}=8\pi G$ \footnote{We use geometrical units $(c=G=1)$ throughout this work.}. From this action Einstein's field equations are obtained as 
\begin{equation}
	\R_{\mu\nu} -\sfrac{1}{2}\,\R \,g_{\mu\nu} = \Mpl^{-2} \, \T_{\mu\nu}
\label{eq:eom_metric}
\end{equation}
where the energy-momentum tensor is given by
\begin{equation}
	\T_{\mu\nu} = \sfrac{\Mpl^2}{\Mpl^2 + \xi \phi^2} \bigg[ \nb_{\!\mu} \, \phi \, \nb_{\!\nu} \, \phi - \sfrac{1}{2} \, g_{\mu\nu} \, \nb^\sg \phi \, \nb_{\!\!\sg} \, \phi + \xi \big( \nb_{\!\mu} \nb_{\!\nu} - g_{\mu\nu} \boxempty \! \big) \phi^2 - g_{\mu\nu}\,V(\phi) \bigg] \:.
\label{eq:stress_tensor}
\end{equation}
On the other hand, equation of motion for the scalar field yields 
\begin{equation}
	\boxempty \! \phi + \xi \phi \R - V'(\phi) = 0
\label{eq:eom_scalar}
\end{equation}
where prime denotes derivative with respect to the scalar field.

We use the following metric
\begin{equation}
    \d s^2 = -\d t^2 + a^2(t) \big( \dt_{ij} + h_{ij} \big) \d x^i \d x^j
\label{eq:metric}
\end{equation}
where $a(t)$ is the scale factor, and we denote $h_{ij}(t,\mf{x})$ as transverse-traceless (TT) part of the metric perturbations. We will neglect the back reaction of the tensor perturbations on the scalar field and the scale factor evolution and, therefore, we get the Friedmann equations of homogeneous background as
\begin{align}
    \left(\sfrac{\dot{a}}{a}\right)^{\!\!2} & = \sfrac{1}{\Mpl^2 + \xi \phi^2} \bigg[ \sfrac{1}{6} \dot{\phi}^2 + \sfrac{1}{6a^2} \big(\Vec{\nb}\phi\big)^2 + \sfrac{1}{3} V(\phi) + \sfrac{\xi}{3a^2} \nb^2 \phi^2 - 2 \xi \sfrac{\dot{a}}{a} \phi \dot{\phi} \bigg] \:, \label{eq:JF_Friedmann} \\[2mm]
    \sfrac{\Ddot{a}}{a} & = \sfrac{-1}{\Mpl^2 + \xi \phi^2} \bigg[ \left(\sfrac{1}{3}+\xi\right)\dot{\phi}^2 - \sfrac{1}{3} V(\phi) + \xi \phi\ddot{\phi} - \sfrac{\xi}{6a^2} \nb^2 \phi^2 + \xi \sfrac{\dot{a}}{a} \phi \dot{\phi} \bigg] \:, \label{eq:JF_Acceleration}
\end{align}
where all quantities of the scalar field should be understood spatially averaged. 

Equation of motion for the scalar field can be recast as
\begin{align}
    \ddot{\phi} + 3\sfrac{\dot{a}}{a}\dot{\phi} - \sfrac{1}{a^2} \nb^2 \phi + 6 \xi \left[ \sfrac{\Ddot{a}}{a} + \left(\sfrac{\dot{a}}{a}\right)^{\!\!2} \right] \phi + \derp{V}{\phi} = 0
\label{eq:JF_Scalar}
\end{align}
with the help of the metric function given in Eq.\ \autoref{eq:metric}.

Following the method described in Ref.\ \cite{Figueroa2007} for the tensor perturbations we have 
\begin{equation}
    \Ddot{u}_{ij} + 3\sfrac{\dot{a}}{a} \dot{u}_{ij} - \sfrac{1}{a^2} \nb^2 u_{ij} = \sfrac{2 \Mpl^{-2}}{a^2} \, \T_{ij}
\label{eq:aux_tensor}
\end{equation}
where the auxiliary tensor, $u_{ij}$, is related to TT part of the tensor perturbations via the following projection operator
\begin{equation}
    h_{ij}(t,\mf{k}) = \Lambda_{ij,lm}(\mf{\hat{k}}) \, u_{lm}(t,\mf{k})
\end{equation}
where $u_{lm}(t,\mf{k})$ is the Fourier transform of $u_{ij}(t,\mf{x})$ and the projection operator is defined as
\begin{equation}
    \Lambda_{ij,lm}(\mf{\hat{k}}) = P_{im}(\mf{\hat{k}}) P_{jl}(\mf{\hat{k}}) - \sfrac{1}{2} P_{ij}(\mf{\hat{k}}) P_{lm}(\mf{\hat{k}}) \:, \qquad P_{ij}(\mf{\hat{k}}) = \dt_{ij} - \sfrac{k_i k_j}{k^2} \:.
\end{equation}
Finally, using Eq.\ \autoref{eq:stress_tensor} in Eq.\ \autoref{eq:aux_tensor} and rearranging the resulting terms we obtain
\begin{equation}
    \Ddot{u}_{ij} + \bigg(3\sfrac{\dot{a}}{a} + \sfrac{2 \xi \phi \dot{\phi}}{\Mpl^2 + \xi \phi^2}\bigg) \dot{u}_{ij} - \sfrac{1}{a^2} \nb^2 u_{ij} = \sfrac{2}{a^2} \sfrac{(1-2\xi) \, \p_i  \phi \, \p_j  \phi - 2 \xi \phi \, \p_i  \p_j  \phi}{\big( \Mpl^2 + \xi \phi^2 \big)} \:.
\end{equation}

We choose to implement the potential for the scalar field same with the one used in Ref.\ \cite{Cutting2018} in the following form 
\begin{equation}
    V(\phi) = \sfrac{1}{2} M^2 \phi^2 + \sfrac{1}{3} \dt \phi^3 + \sfrac{1}{4} \lb \phi^4
\label{eq:potential}
\end{equation}
where $M$, $\dt$, and $\lb$ are constants. This form is enough for our purposes and it will provide an easy and direct comparison of the outcomes.

\section{THE EFFECTIVE POTENTIAL\label{sec:effective_potential}}

\begin{figure*}[!t]
	\centering

	\begin{tabular}{@{}c@{}}\hspace{-5mm}
		\includegraphics[width=.56\linewidth]{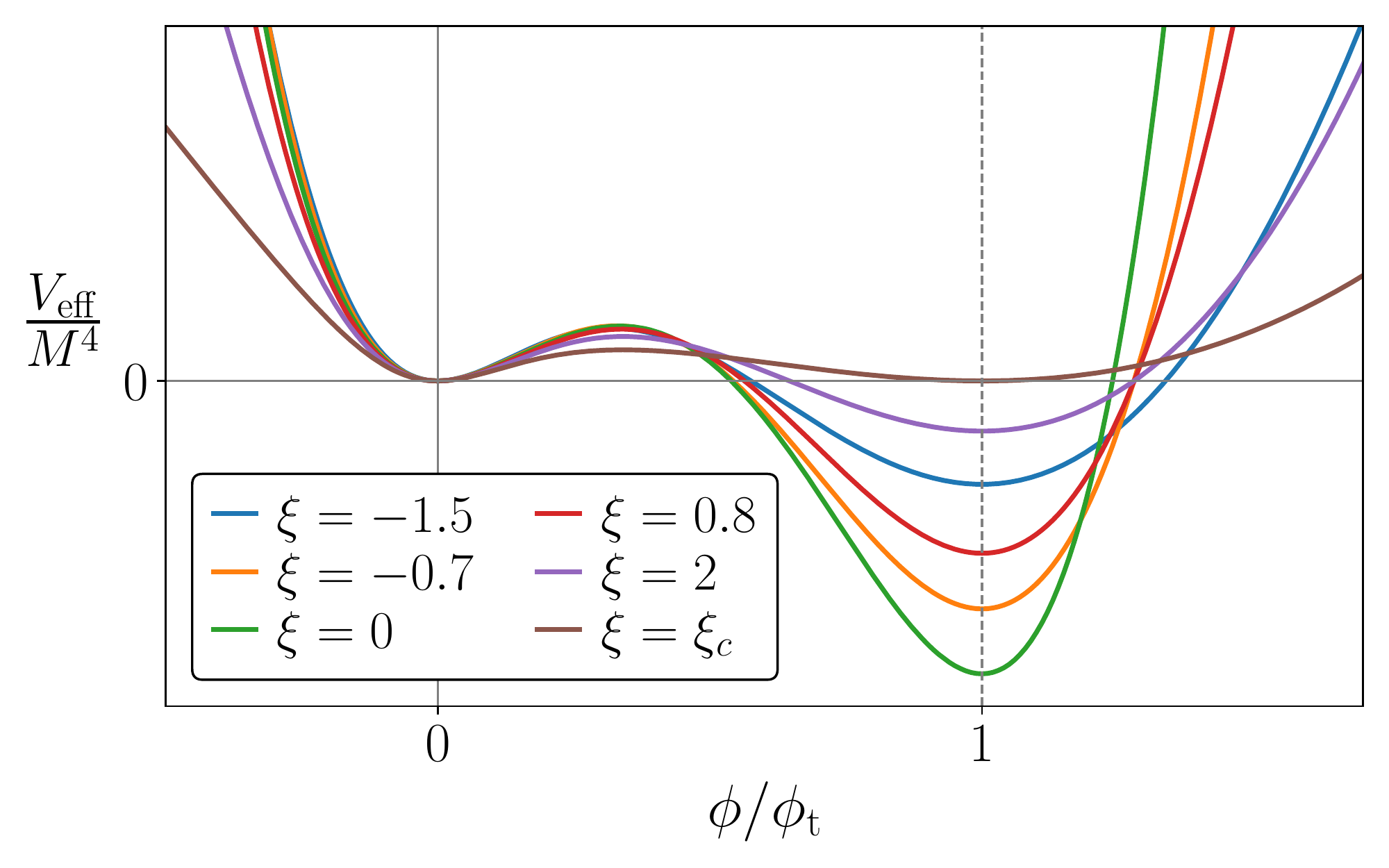}
	\end{tabular} \hspace{3mm}
	\begin{tabular}{@{}c@{}}
		\includegraphics[width=.41\linewidth]{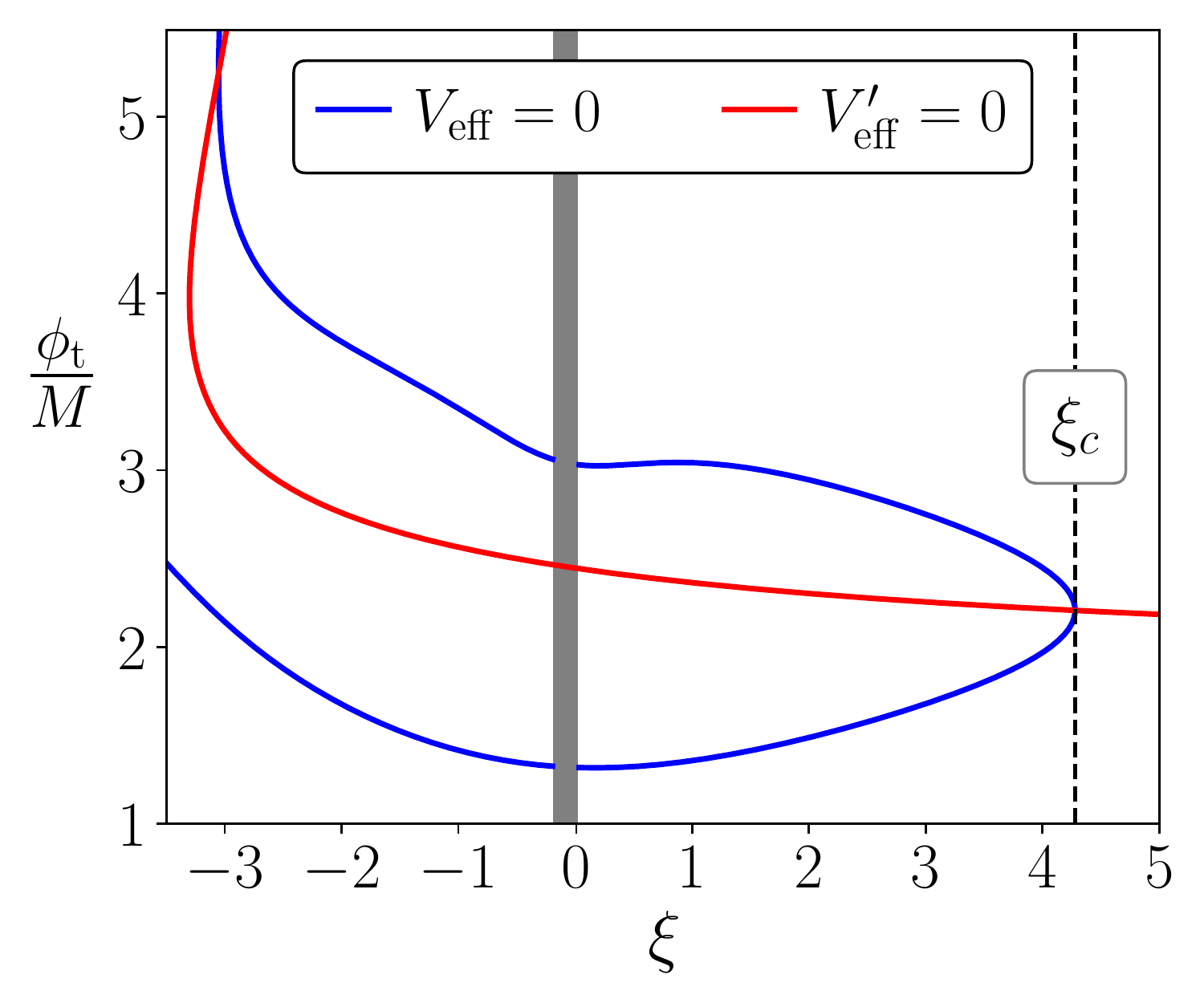}
	\end{tabular}
	
	\vspace{-2mm}
	
	\caption{Here the potential parameters are $\dt/M=-1.632$ and $\lb=0.5$. (LEFT) The effective potential given in Eq.\ \autoref{eq:effective_potential} for different values of the coupling constant, $\xi$. $\phi_{\rm{t}}$ indicates the value of the scalar field in true vacuum. (RIGHT) Two curves corresponding to $V_{\rm{eff}}=0$ and $V'_{\rm{eff}}=0$ which intersect at critical $\xi$, i.e. $\xi_c \approx 4.28$ at which two minima becomes degenerate. Vertical gray region indicates the forbidden region, $-1/6 \leq \xi_{\rm{forbidden}} \leq 0$, where the effective potential becomes discontinuous.}
\label{fig:effective_potential} \vspace{0.2cm}
\end{figure*}

In this section we try to understand the effect of non-minimal coupling on the dynamics of the phase transition by investigating the behavior and, possibly, the validity range of the coupling constant, $\xi$, via an effective potential. Therefore, this analysis will give a possible restriction on the free parameter of the model and explicitly show the impact of a chosen value that will provide a guide for the simulations. To this end, we rearrange Eq.\ \autoref{eq:JF_Scalar} with the help of Eqs.\ \autoref{eq:JF_Friedmann} and \autoref{eq:JF_Acceleration} and get the equation of motion for the scalar field in the following form
\begin{equation}
    \ddot{\phi} + 3 \sfrac{\dot{a}}{a} U_1(\phi) \dot{\phi} + U_2(\phi) \dot{\phi}^2 - \big[U_1(\phi)-U_2(\phi)\big] \sfrac{1}{a^2} \big(\nb \phi\big)^2 +  \derp{V_{\rm{eff}}}{\phi} = 0 
\end{equation}
where we have defined
\begin{equation}
    U_1(\phi) \equiv \sfrac{\Mpl^2 + 3\xi\phi^2}{\Mpl^2 + \xi(1+6\xi)\phi^2} \;, \qquad\qquad U_2(\phi) \equiv \sfrac{\xi(1+6\xi)}{\Mpl^2 + \xi(1+6\xi)\phi^2}
\end{equation}
and the derivative of the effective potential is given by
\begin{equation}
    \derp{V_{\rm{eff}}}{\phi} = \sfrac{(\Mpl^2+\xi\phi^2) \derp{V}{\phi} - 4\xi\phi V(\phi) }{\Mpl^2 + \xi (1 + 6\xi)\phi^2} \:.
\end{equation}
Now, by plugging the potential given in Eq.\ \autoref{eq:potential} into this expression we obtain
\begin{equation}
    \derp{V_{\rm{eff}}}{\phi} = \sfrac{\Mpl^2 M^2 \phi + \Mpl^2 \dt \phi^2 - (\xi M^2 - \Mpl^2 \lb)\phi^3 - \sfrac{1}{3}\xi\dt\phi^4}{\Mpl^2 + \xi (1 + 6\xi)\phi^2}
\end{equation}
integration of which yields
\begin{align}
    V_{\rm eff}(\phi) = & \,V_0 - A + B \phi - C \phi^2 - D \phi^3 + E_1 \ln\!\big( 1 + E_2 \, \phi^2 \big) - F \tan^{-1} \!\big( \sqrt{E_2} \, \phi \big)
\label{eq:effective_potential}
\end{align}
where
\begin{equation}
\begin{aligned}
    &A = \sfrac{\Mpl^2 C}{K}  \;, \qquad B = \sfrac{\Mpl^2 \dt (\xi + 3K)}{3K^2} \;, \qquad C = \sfrac{M^2\xi-\Mpl^2\lb}{2K} \;, \qquad D = \sfrac{\dt\xi}{9K} \\[2mm]
    &E_1 = \sfrac{\Mpl^2\big( M^2 + 2 C \big)}{2K} \;, \qquad E_2 = \sfrac{K}{\Mpl^2} \;, \qquad F = \sfrac{B \Mpl^2}{\sqrt{K}} \;, \qquad K = \xi(1+6\xi)
\end{aligned}
\label{eq:eff_pot_params}
\end{equation}
and $V_0$ is the integration constant which we set to $A$. Although this effective potential is unbounded from one side and it can be fixed with the addition of some higher order correction terms to the potential given in Eq.\ \autoref{eq:potential}, we do not prefer to do that since the current form does not have any negative effects on solving the classical equations of motion and, therefore, it is sufficient for our purposes. On the other hand, Eq.\ \autoref{eq:eff_pot_params} puts a mathematical constraint on the coupling constant to make the effective potential analytical and that is $\xi(1+6\xi)\geq0$ which leads to a forbidden region in parameter space as $-1/6 \leq \xi_{\rm{forbidden}} \leq 0$.

We give an example plot of the effective potential in Fig.\ \autoref{fig:effective_potential} for some fixed parameter values of $\dt/M$ and $\lb$. As seen from the figure the difference between true and false vacuums gets smaller when the value of the coupling constant increases which means that the bubble walls get thinner. In addition to that, there is a value, $\xi=\xi_c$, at which two minima becomes degenerate. For values bigger than that first-order transition does not occur or may take place in the opposite way. The latter possibility was discussed in detail in Refs.\ \cite{Lee2006,Lee2012}.

\section{THE BUBBLE PROFILE\label{sec:bubbleProfile}}

Bounce solutions for the non-minimally coupled scalar field were studied previously in different contexts \cite{Lee2006,Salvio2016,Czerwinska2016,Rajantie2016}. Here, we also examine the bubble profiles by solving the equations numerically in order to compare them with the thin-wall approximation \cite{Coleman1977_1} and determine whether it is appropriate to implement here or not.

As mentioned earlier, in flat space-time the most probable bubble configuration was found to be $O(4)$ symmetric configuration \cite{Coleman1978}. Although there is no such proof for a curved space-time to this day, in this work we assume that the same $O(4)$ symmetrical bubble profile is valid for both the scalar field and the metric tensor. Therefore, the most general rotationally invariant Euclidean metric can be written as
\begin{equation}
    \d s^2 = \d \eta^2 + \rho^2(\eta) \big[ \d \chi^2 + \sin^2\!\chi \big(\d \theta^2 + \sin^2\!\theta \, \d \varphi^2 \big) \big]
\end{equation}
which gives rise to the following Euclidean equations of motion
\begin{align}
    & \phi'' + \sfrac{3\rho'}{\rho} \phi' + \sfrac{\xi (1 + 6 \xi)\phi}{\Mpl^2 + \xi (1 + 6\xi)\phi^2} \phi'^2 - \derp{V_{\rm{eff}}}{\phi} = 0 \:, \label{eq:eom_scalar_eucl} \\[2mm]
    & \rho'^2 = 1 + \sfrac{\rho^2}{3\big(\Mpl^2+\xi\phi^2\big)} \bigg[ \sfrac{1}{2} \phi'^2 - V(\phi) - 6\xi \phi \phi' \sfrac{\rho'}{\rho} \bigg] \:, \label{eq:eom_scalar_rho1}  \\[2mm]
    & \rho'' = - \sfrac{\rho}{3\big(\Mpl^2+\xi\phi^2\big)} \bigg[ \phi'^2 + V(\phi) + 3\xi \big( \phi'^2 + \phi \phi'' + \phi \phi' \sfrac{\rho'}{\rho} \big) \bigg] \label{eq:eom_scalar_rho2} \:,
\end{align}
for the scalar field and the Einstein's field equations where prime designates the derivative with respect to $\eta$. In order to find the bounce solutions we solve Eqs.\ \autoref{eq:eom_scalar_eucl} and \autoref{eq:eom_scalar_rho2} numerically with the shooting method and use Eq.\ \autoref{eq:eom_scalar_rho1} as the constraint. We impose the boundary conditions as $\phi(0) = \phi_{\rm{bs}}$, $\phi'(0) = 0$, $\rho(0) = 0$, and $\rho'(0) = 0$ where $\phi_{\rm{bs}}$ is the critical value that provides $\phi(\eta \rightarrow \pm \infty) = 0$.

\begin{figure*}[!t]
	\centering

	\begin{tabular}{@{}c@{}}\hspace{-5mm}
		\includegraphics[width=.49\linewidth]{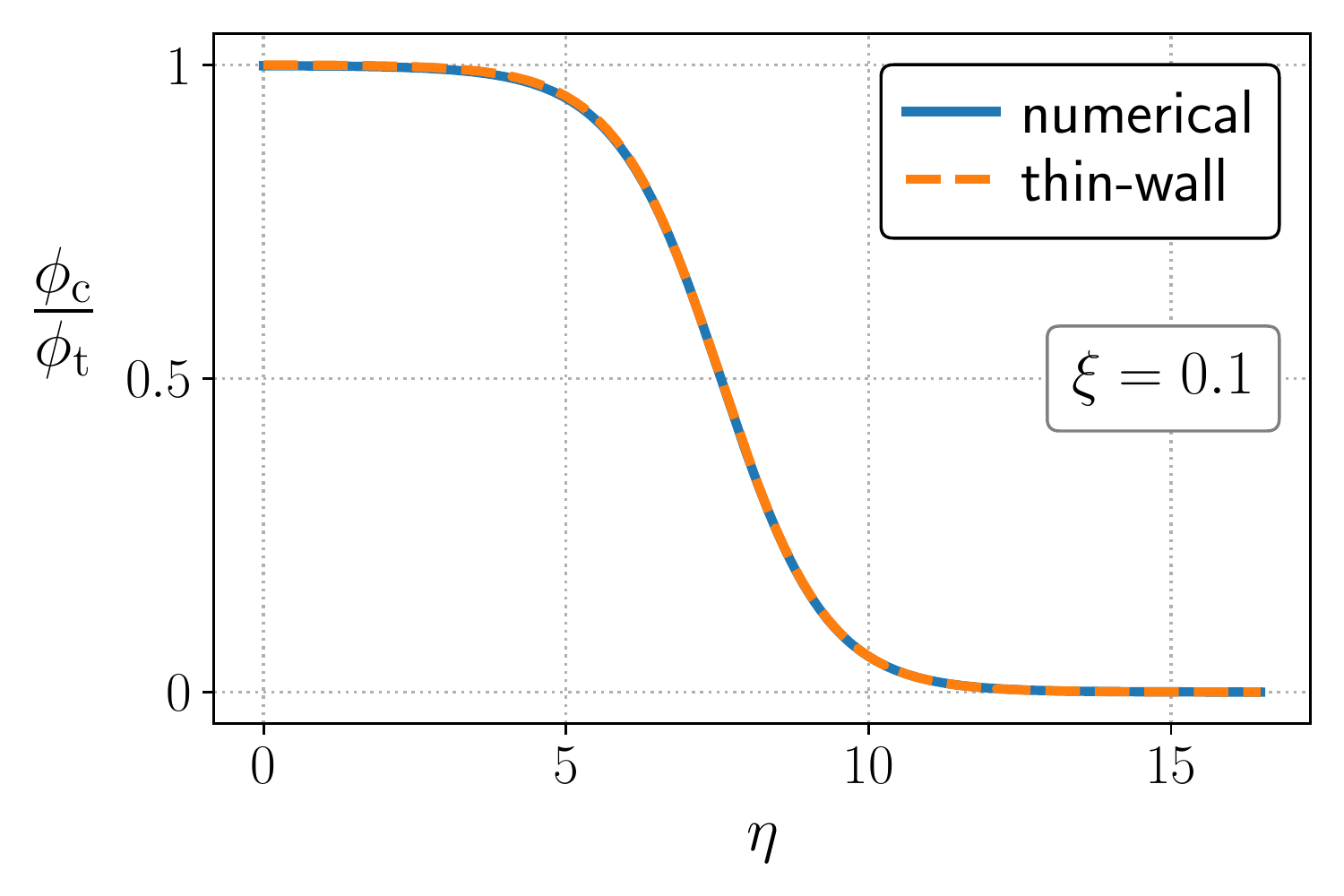}
	\end{tabular}
	\begin{tabular}{@{}c@{}}
		\includegraphics[width=.49\linewidth]{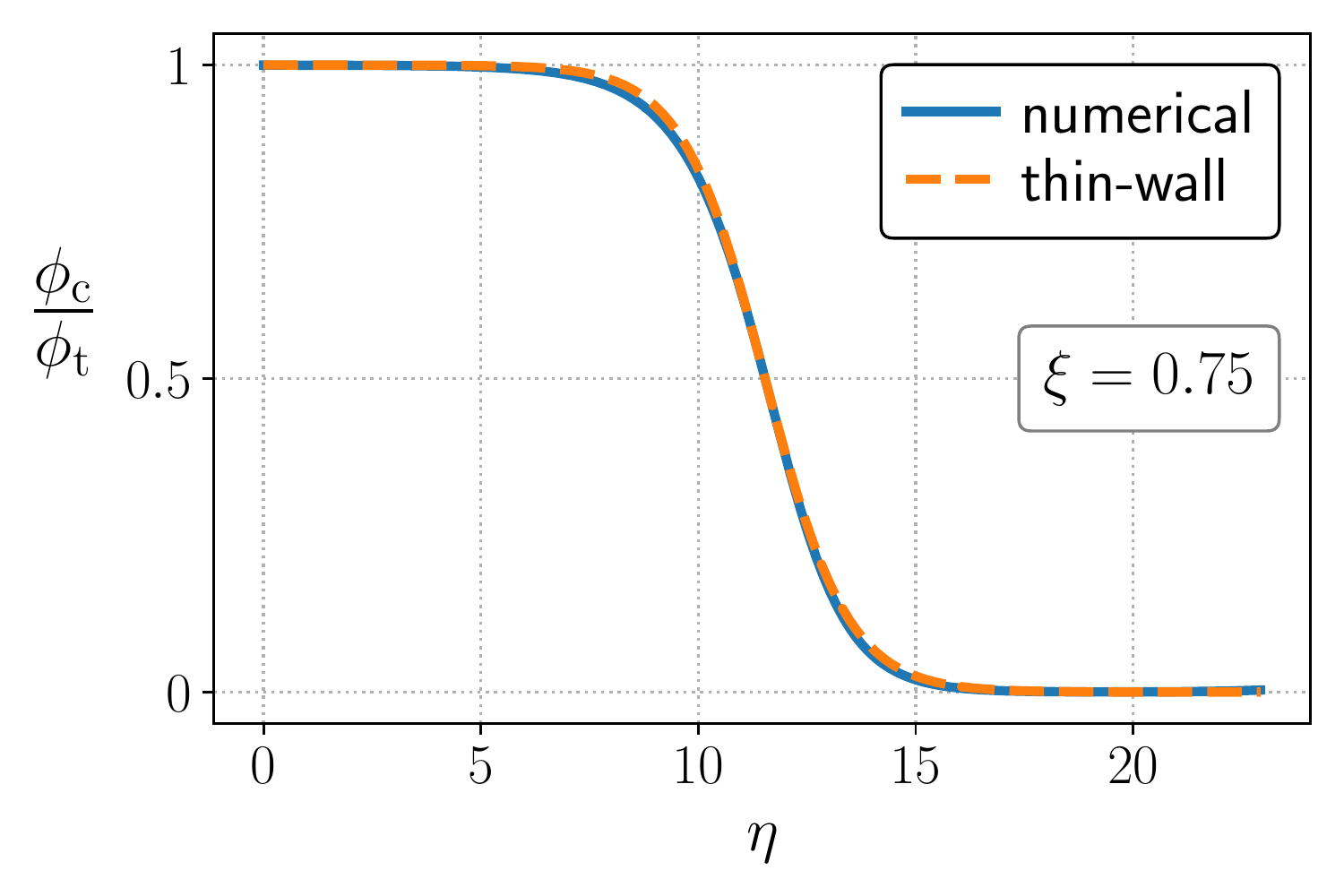}
	\end{tabular}
	
	\vspace{-2mm}
	
	\caption{Critical bubble profiles for two different $\xi$ values with $\dt/M = -1.632$ and $\lb=0.5$. The blue solid lines represent numerical bounce solution obtained by solving Eqs.\ \autoref{eq:eom_scalar_eucl} and \autoref{eq:eom_scalar_rho2} whereas the dashed orange lines stand for the thin-wall approximation given in Eq.\ \autoref{eq:thin-wall}.}
\label{fig:bubble_profiles} \vspace{0.5cm}
\end{figure*}

We give two examples of the critical bubble profiles in Fig.\ \autoref{fig:bubble_profiles} for the same parameter values used to produce the results in Fig.\ \autoref{fig:effective_potential}. In the same figures we also plot the graphs of the equation for the thin-wall approach as
\begin{equation}
    \phi(r) = \sfrac{\phi_{\rm{t}}}{2} \bigg[ 1 - \tanh \! \bigg( \sfrac{r-R_{\rm{c}}}{l_0} \bigg) \bigg]
\label{eq:thin-wall}
\end{equation}
where $R_{\rm{c}}$ and $l_0$ are the critical radius and the bubble wall length, respectively, which can be found from the following expressions \cite{Cutting2018}
\begin{equation}
    \phi(R_{\rm{c}}) = \sfrac{1}{2} \phi_{\rm{t}} \:, \qquad \phi(r^\pm) = \sfrac{\phi_{\rm{t}}}{2} \bigg[ 1 - \tanh \! \bigg( \!\! \pm \sfrac{1}{2} \bigg) \bigg] \:, \qquad l_0 = r^+ - r^- \:.
\end{equation}

As a result of this brief analysis, we can conclude that the thin-wall approximation can be safely applied for at least $\xi<1$ which already describes much bigger region than the one that we are interested in this work. Therefore, we use Eq.\ \autoref{eq:thin-wall} for the initial bubble profiles at the start of our simulations.

We also calculate the parameters for different values of the coupling constant as given in Table \autoref{tab:bubble_profile}. To this end, we use $\lb=0.5$ for all solutions together with $\dt/M=-1.632$ and $\dt/M=-1.56$. As seen from the table, the critical radius and the wall thickness increase as value of the coupling constant raises while value of the scalar field in true vacuum and the vacuum energy density decrease. Since the vacuum energy density is defined as the difference between two minima of the potential, this result confirms the illustration given in Fig.\ \autoref{fig:effective_potential}. In addition to that, the ratio $R_{\rm{c}}/l_0$ grows as the coupling constant increases which means the thin-wall limit is getting more applicable but probably with a slightly different form than Eq.\ \autoref{eq:thin-wall}.

\begin{table}[t]
\renewcommand{\arraystretch}{0.85}
\begin{tabular}{|C{1cm}|C{1.5cm}|C{1.2cm}||C{1.2cm}|C{1.3cm}|C{1.3cm}|C{1.3cm}|C{1.7cm}|} \hline
$\lb$ & $\dt/M$ & $\xi$ & $R_c M$ & $l_0 M$ & $R_c/l_0$ & $\phi_{\rm{t}}/M$ & $\rho_{\rm{vac}}/M^4$ \\ \hline\hline
\multirow{9}{*}{0.5} & \multirow{6}{*}{-1.632}  & 0 & 7.15 & 1.71 & 4.18 & 2.45 & 0.495 \\ 
                    &                           & 0.1 & 7.56 & 1.74 & 4.35 & 2.44 & 0.482 \\ 
                    &                           & 0.25 & 8.14 & 1.75 & 4.65 & 2.42 & 0.450 \\ 
                    &                           & 0.5 & 9.47 & 1.81 & 5.23 & 2.40 & 0.379 \\ 
                    &                           & 0.75 & 11.54 & 1.90 & 6.07 & 2.38 & 0.305 \\ 
                  &                             & 1 & 15.25 & 2.01 & 7.59 & 2.36 & 0.240 \\ \cline{2-8} 
                  & \multirow{3}{*}{-1.56}      & 0 & 14.3 & 1.83 & 7.81 & 2.22 & 0.189 \\ 
                  &                             & 0.25 & 17.52 & 1.87 & 9.37 & 2.21 & 0.176 \\ 
                  &                             & 0.5 & 24.11 & 1.90 & 12.69 & 2.20 & 0.148 \\ \hline
\end{tabular}
\caption{Parameter values used in the simulations. $\lb$ and $\dt/M$ are the same values with Ref.\ \cite{Cutting2018} in order to compare the results and to see the effect of various $\xi$ values. Critical radius ($R_c$), bubble wall thickness ($l_0$), scalar field value in true vacuum ($\phi_{\rm{t}}$), and the vacuum energy density ($\rho_{\rm{vac}}$) are calculated as explained in the text.}
\label{tab:bubble_profile}
\end{table}

\section{THE MODEL FOR SIMULATIONS\label{sec:modelForSimulations}}

As discussed in earlier sections, we neglect the contribution of the scale factor in equation for the scalar field and solve the following form
\begin{equation}
    \ddot{\phi} - \nb^2 \phi + \derp{V}{\phi} = 0 \:,
\end{equation}
which, actually, is the case of the minimal coupling. However, the equation for the tensor perturbations turns into
\begin{equation}
    \Ddot{u}_{ij} + \sfrac{2 \xi \phi \dot{\phi}}{\Mpl^2 + \xi \phi^2} \dot{u}_{ij} - \nb^2 u_{ij} = 2 \, \sfrac{(1-2\xi) \, \p_i  \phi \, \p_j  \phi - 2 \xi \phi \, \p_i  \p_j  \phi}{\Mpl^2 + \xi \phi^2}
\end{equation}
that still has additional source terms on right-hand side and a friction term on the left-hand side both depending on $\xi$. Therefore, we expect some differences even if we neglect the expansion of the Universe. Since we do not take into account the change in the equation of motion for the scalar field due to the non-minimal coupling, we will use small values for the coupling constant to make sure that the same simulation parameters are still valid approximately and, furthermore, to ensure compatibility with the constraints found in the literature as we will mention shortly.

Finally, we calculate the GW energy density through
\begin{equation}
	\rho_{\rm gw}(\mf{x},t) = \sfrac{1}{32\pi G} \sum_{i,j} \dot{h}_{ij}(\mf{x},t) \, \dot{h}_{ij}(\mf{x},t)
\label{eq:gw_density}
\end{equation}
where dots represent the derivative with respect to time.

\section{NUMERICAL METHODS AND THE POWER SPECTRA\label{sec:numerical_methods}}
Our code has been built on \texttt{Python} programming language and to speed it up on some intensive iterations, such as calculation of power spectra and projection operations on Fourier space, \texttt{Cython} \cite{Cython} extension has been used. The code works parallel based on pencil decomposition and communication between processes is provided by \texttt{mpi4py} \cite{mpi4py} package. We constructed similar algorithms given in Ref.\ \cite{Mortensen2016} for the Fourier transforms in parallel computing in \texttt{Python} and with the guidance of the same study we wrote a routine for the inverse Fourier transforms that is necessary to calculate the spatial distribution of GW energy density. The implemented method for solving the differential equations is the staggered leapfrog together with 7-point stencil for the Laplacian operator. 

The equation set we use for the computation is
\begin{align}
    & \psi'' - \nb^2 \psi + \psi + \al\bt \psi^2 + \lb\bt^2 \psi^3 = 0 \:, \label{eq:eom_phi_numerical} \\[2mm]
    & u''_{ij} + \sfrac{2 \xi \psi \psi'}{\mpl^2 / \bt^2 + \xi \psi^2} u'_{ij} - \nb^2 u_{ij} = 2 \, \sfrac{(1-2\xi) \, \p_i  \psi \, \p_j  \psi - 2 \xi \psi \, \p_i  \p_j  \psi}{\mpl^2 / \bt^2 + \xi \psi^2}
\label{eq:eom_uij_numerical}
\end{align}
with the following definitions
\begin{align}
    \psi \equiv \sfrac{\phi}{\phi_{\rm{t}}} \;, \quad \mpl^2 \equiv \sfrac{\Mpl^2}{M^2} \;, \quad \Vec{x} \rightarrow M \Vec{x} \;, \quad \d \tau \equiv M \d t \;, \quad \al \equiv \sfrac{\dt}{M}  \;, \quad  \bt \equiv \sfrac{\phi_{\rm{t}}}{M} 
\label{eq:eom_params_numerical}
\end{align} 
and the prime denotes the derivative with respect to $\tau$. For the initial conditions in time we set $\psi(\tau=0)=\psi_{\rm{init}}$, $\psi'(\tau=0)=0$, and $u_{ij}(\tau=0) = u'_{ij}(\tau=0) = 0$ where $\psi_{\rm{init}}$ represents random initial locations of nucleated bubbles and we use periodic boundary conditions for the spatial part.

Appearance of the first derivative in the second term on the left-hand side of Eq.\ \autoref{eq:eom_uij_numerical} is problematic for our algorithm. To solve this problem we have exploited a method adopted in Ref.\ \cite{LATTICEEASY} to correct the staggered leapfrog for this particular equation. The spatial resolution and the Courant factor have been chosen as $\d x=0.22$ and $c=0.2$, respectively. 

We apply the following definitions of the power spectrum for the scalar field
\begin{equation}
	\sfrac{\P_{\phi}(\mf{k},t)}{\phi_{\rm{t}}^2} = \sfrac{k^3}{2\pi^2} \langle \psi(\mf{k},t) \, \psi^*(\mf{k},t) \rangle
\end{equation}
and for the gravitational waves
\begin{align}
    \der{\Omega_{\rm gw}}{\ln k} & = \sfrac{1}{32 \pi G \rho_c} \sfrac{k^3}{2\pi^2} \langle \dot{h}_{ij}(\mf{k},t) \, \dot{h}_{ij}(\mf{k},t) \rangle \\[2mm]
    & = \sfrac{\Mpl^2}{4 \rho_c} \sfrac{k^3}{2\pi^2} \langle \Lambda_{ij,lm} \dot{u}_{lm}(\mf{k},t) \, \Lambda_{ij,pr} \dot{u}^*_{pr}(\mf{k},t) \rangle
\end{align}
where the angle brackets denote the spatial averaging. However, we implement the following normalization for the graphical representations \cite{Turner1992_norm,Cutting2018}
\begin{equation}
	\der{\Omega_{\rm gw}}{\ln k} \longrightarrow \sfrac{1}{(H_* R_* \Omega_{\rm vac})^2} \der{\Omega_{\rm gw}}{\ln k} 
\end{equation}
where $R_*$ is the mean bubble separation given as $R_* = (\V / N_{\rm{b}})^{1/3}$ in which $\V$ and $N_{\rm{b}}$ are the physical volume of the simulation box and the number of bubbles, respectively. 

Although this does not change the results of our simulations, in order to reduce the effect of the numerical artifacts occurring at ultraviolet frequency regime we partly follow the methods described in Ref.\ \cite{powerSpectraAverage}.

\section{RESULTS\label{sec:results}}

\begin{figure*}[!t]
	\centering

    \begin{tabular}{@{}c@{}}\hspace{-7mm}
		\includegraphics[width=.5\linewidth]{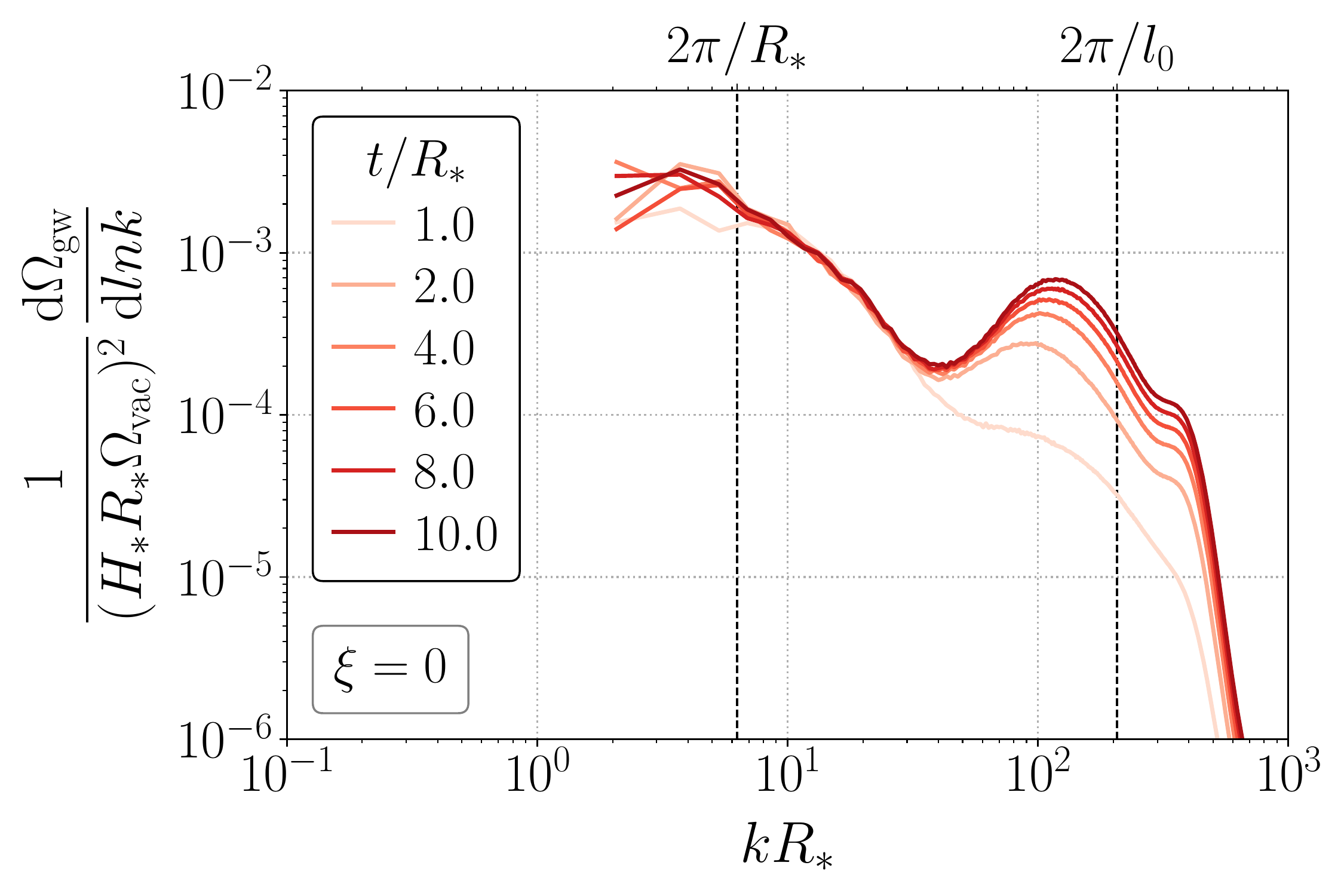}
	\end{tabular}
	\begin{tabular}{@{}c@{}}\hspace{-2mm}
		\includegraphics[width=.5\linewidth]{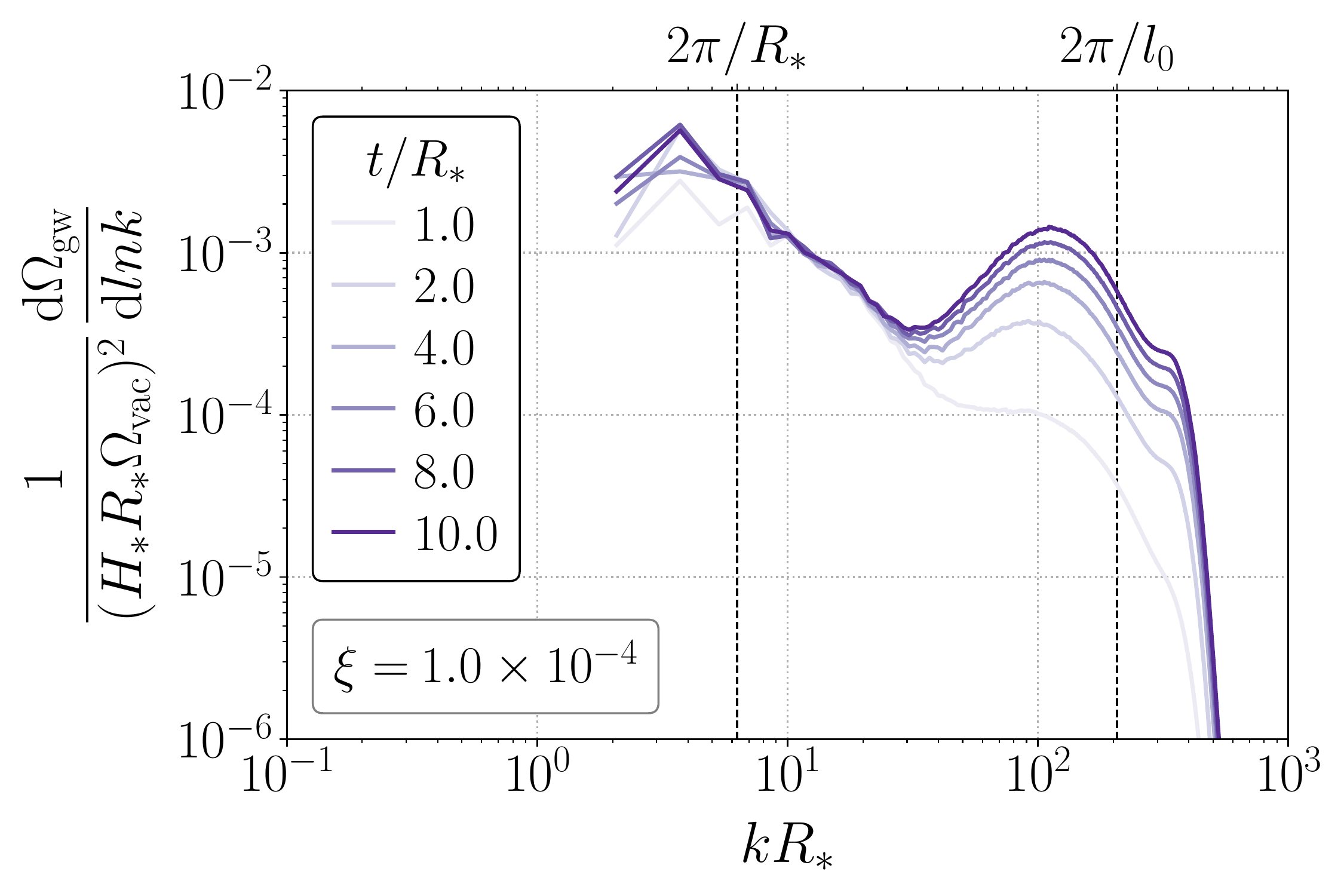}
	\end{tabular}
	
	\vspace{2mm}
	
	\begin{tabular}{@{}c@{}}\hspace{-7mm}
		\includegraphics[width=.5\linewidth]{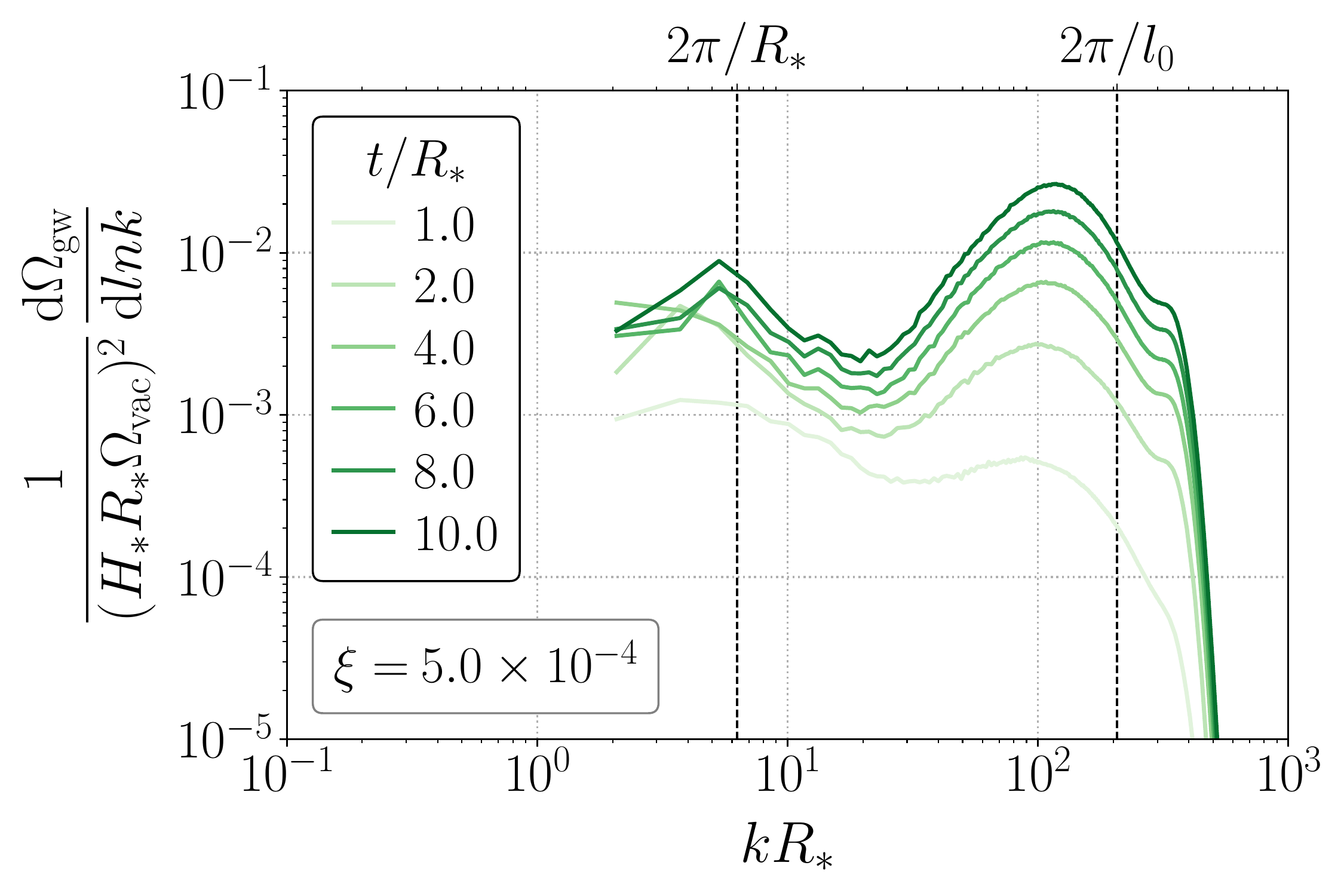}
	\end{tabular}
	\begin{tabular}{@{}c@{}}\hspace{-2mm}
		\includegraphics[width=.5\linewidth]{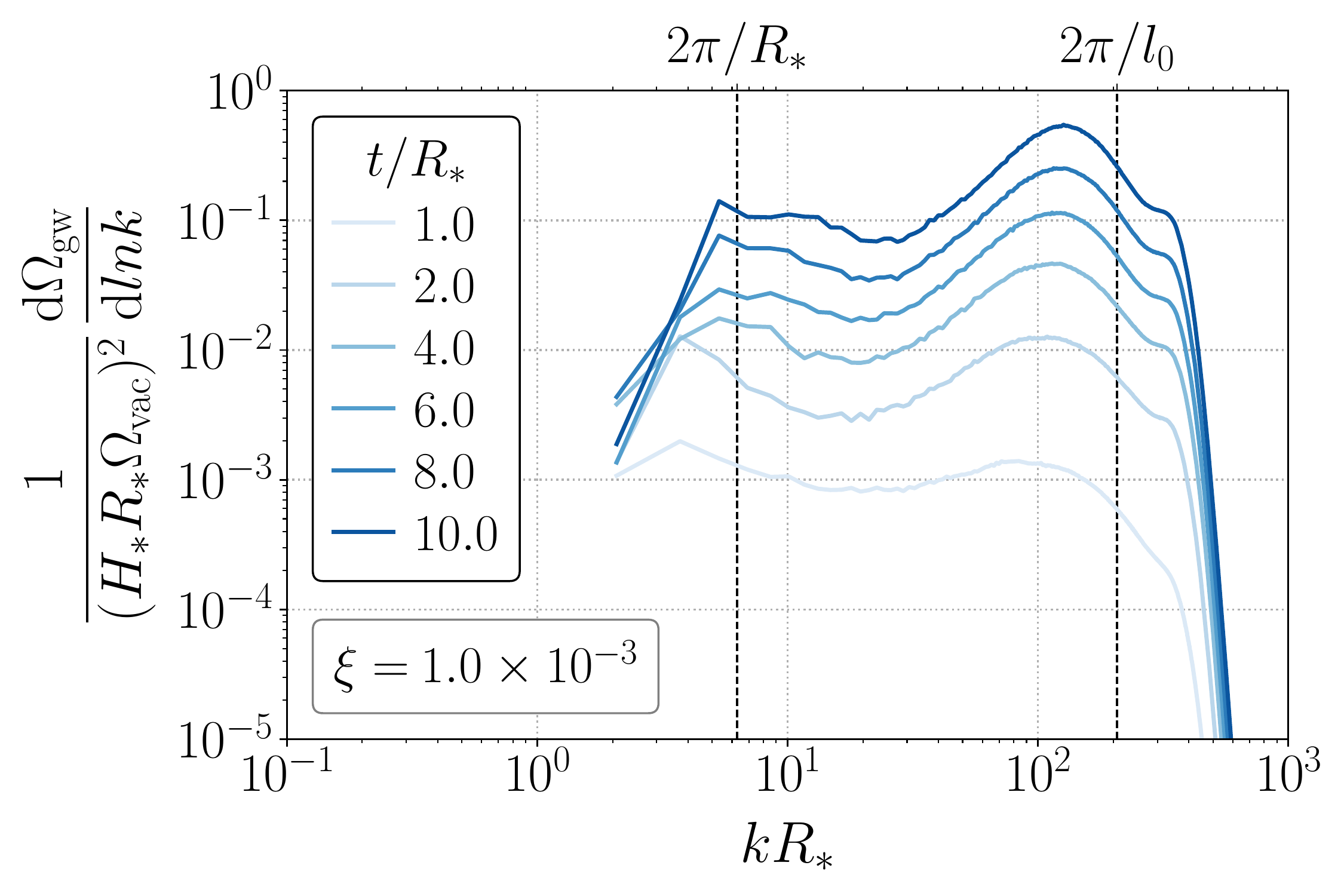}
	\end{tabular}
	
	\vspace{2mm}
	
	\caption{Power spectra of GW energy density for different values of the coupling constant. All simulations have been run in a cubic simulation box of 1024 points per edge with 64 bubbles nucleated simultaneously.}
\label{fig:pspec_gw} 
\end{figure*}

\begin{figure*}[!h]
	\centering

	\begin{tabular}{@{}c@{}}\hspace{-5mm}
		\includegraphics[width=.35\linewidth,trim={0 0 0 15mm},clip]{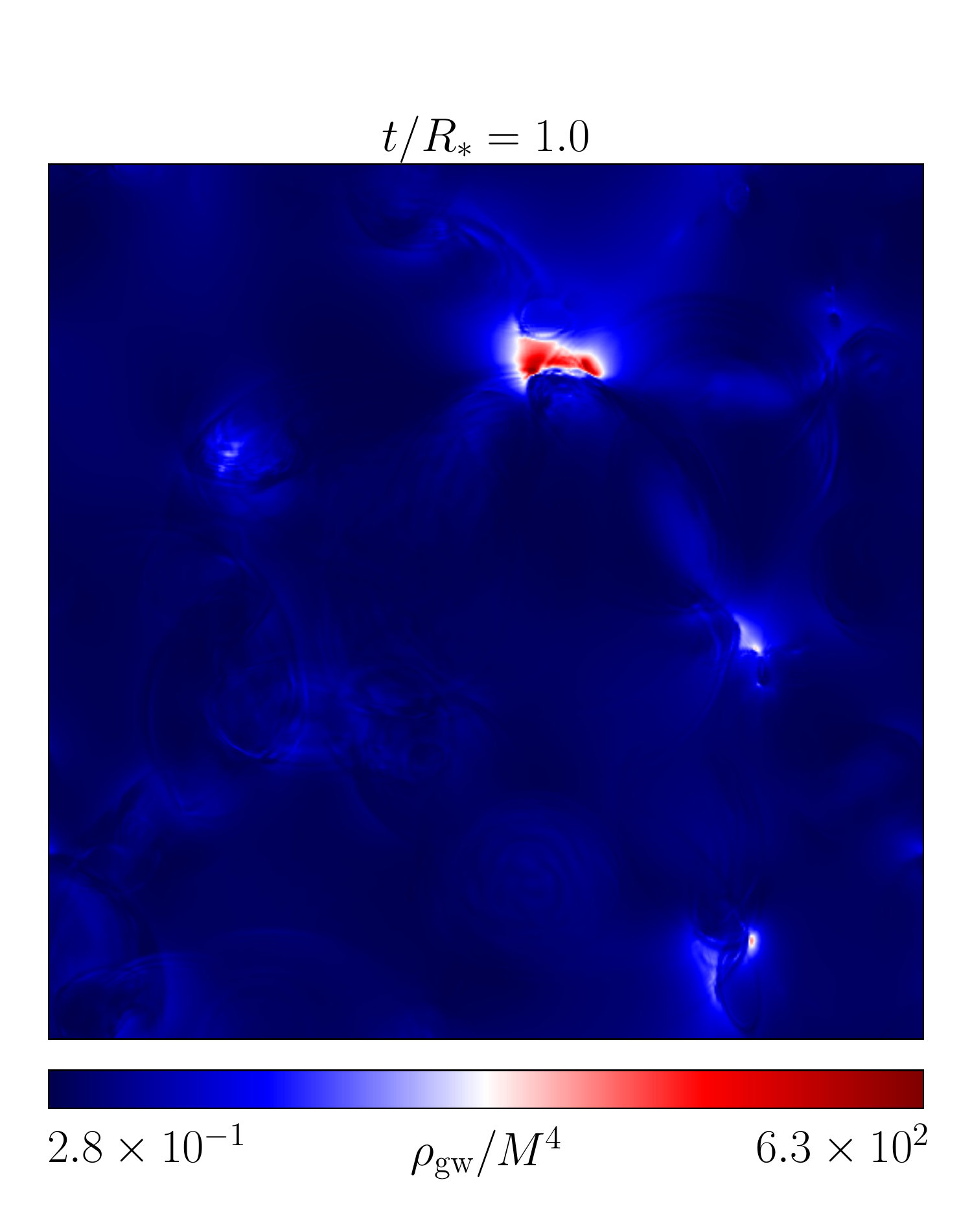}
	\end{tabular}
	\begin{tabular}{@{}c@{}}\hspace{-5mm}
		\includegraphics[width=.35\linewidth,trim={0 0 0 15mm},clip]{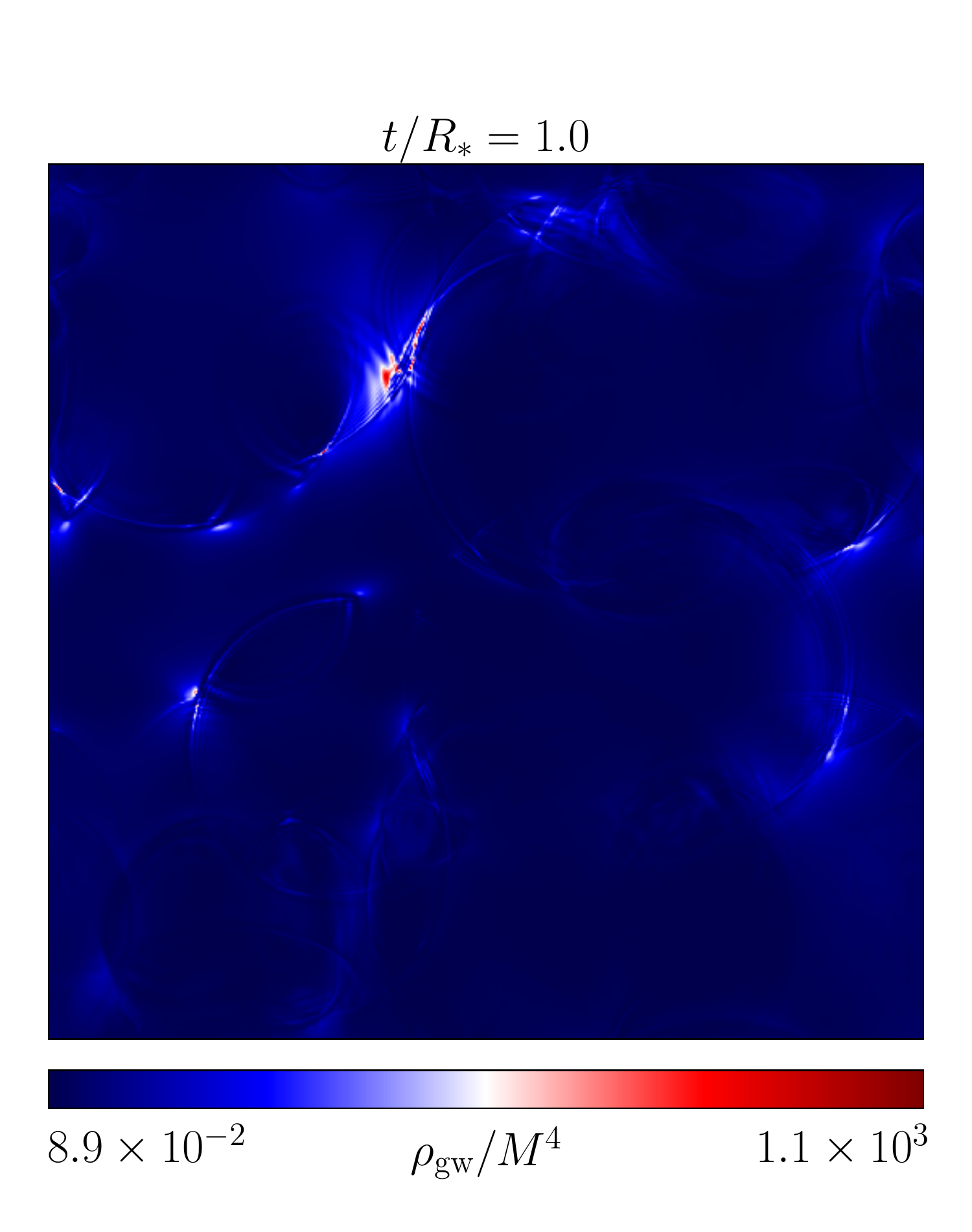}
	\end{tabular}
	\begin{tabular}{@{}c@{}}\hspace{-5mm}
		\includegraphics[width=.35\linewidth,trim={0 0 0 15mm},clip]{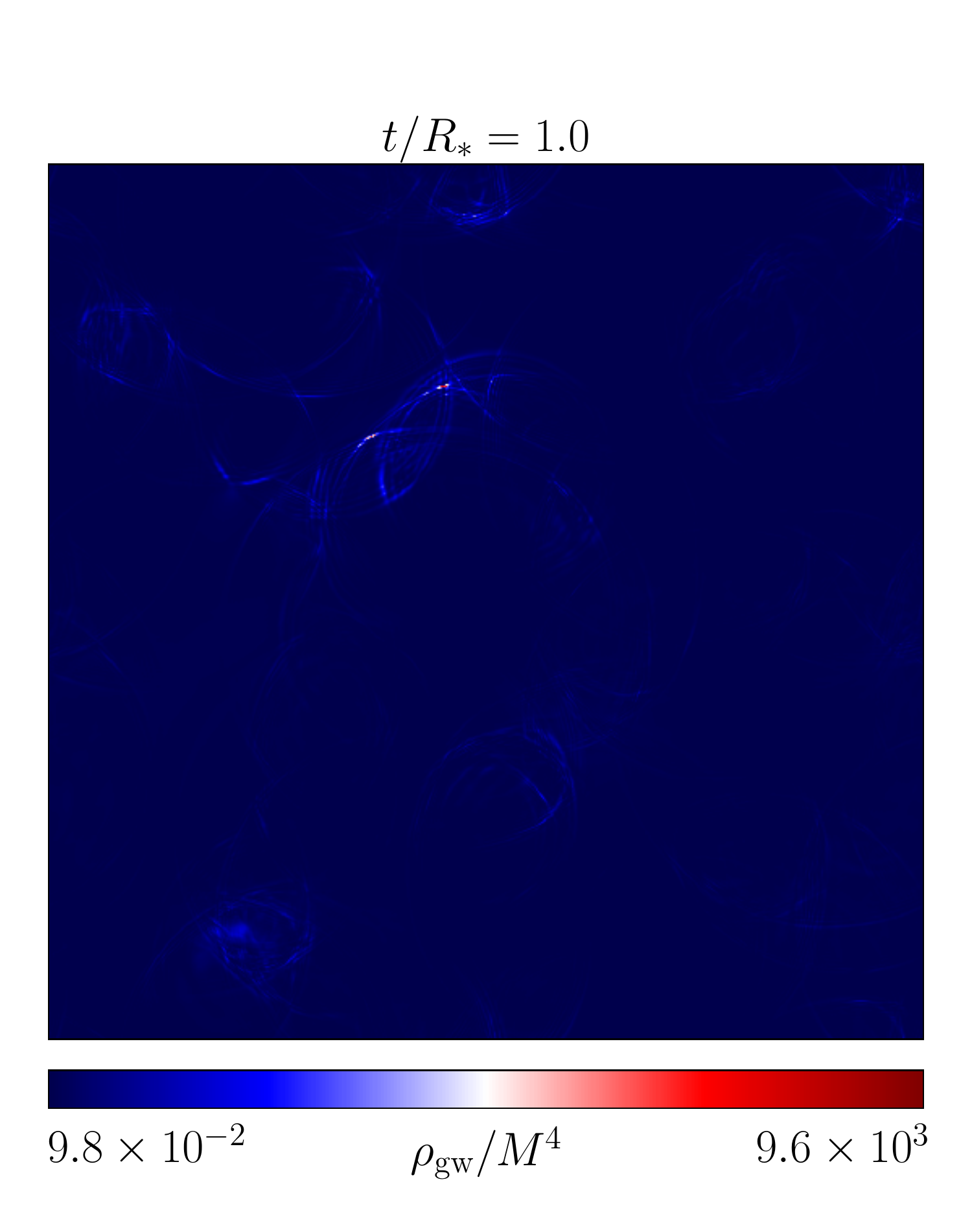}
	\end{tabular}
	
	\vspace{-2mm}
	
	\begin{tabular}{@{}c@{}}\hspace{-5mm}
		\includegraphics[width=.35\linewidth,trim={0 0 0 15mm},clip]{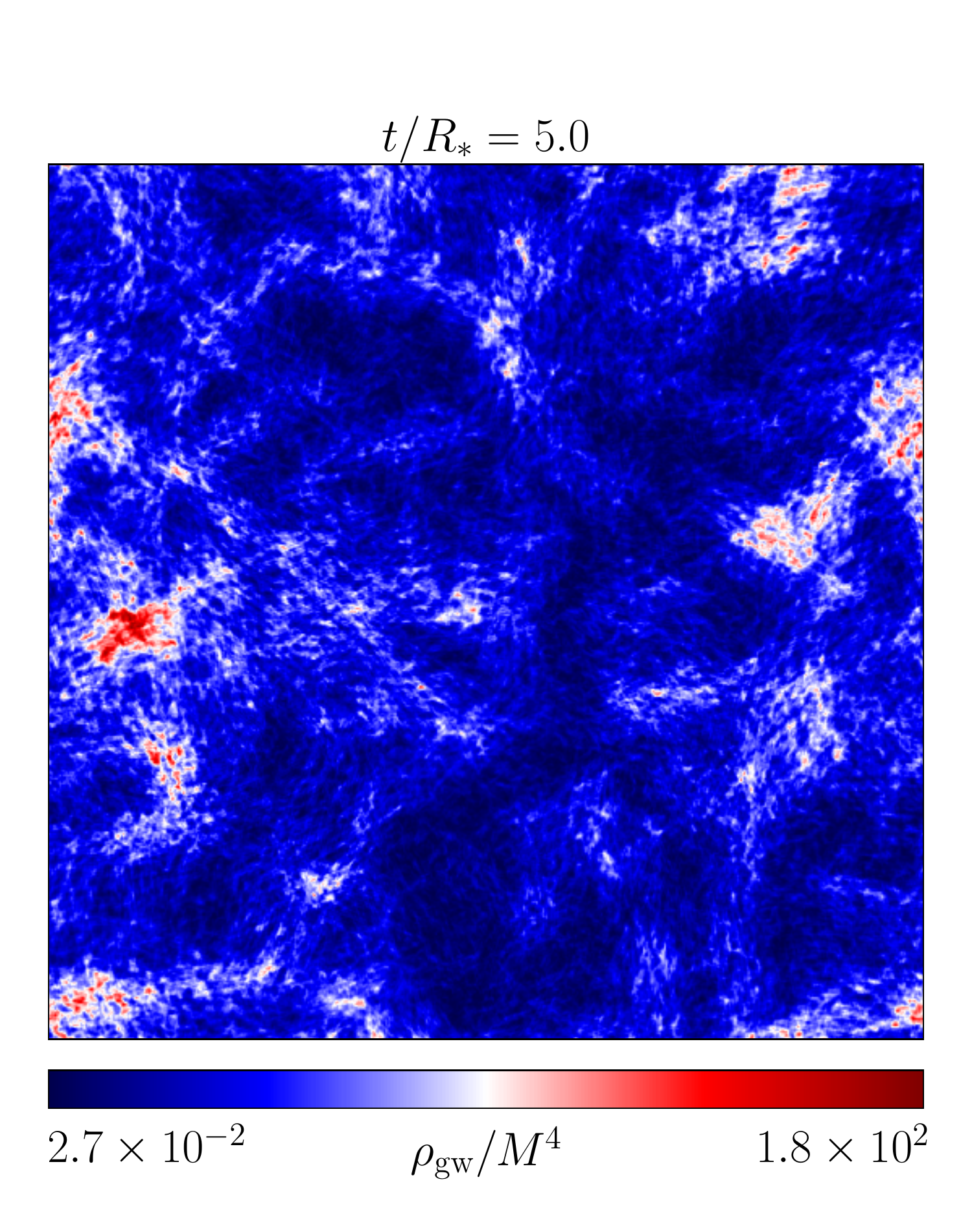}
	\end{tabular}
	\begin{tabular}{@{}c@{}}\hspace{-5mm}
		\includegraphics[width=.35\linewidth,trim={0 0 0 15mm},clip]{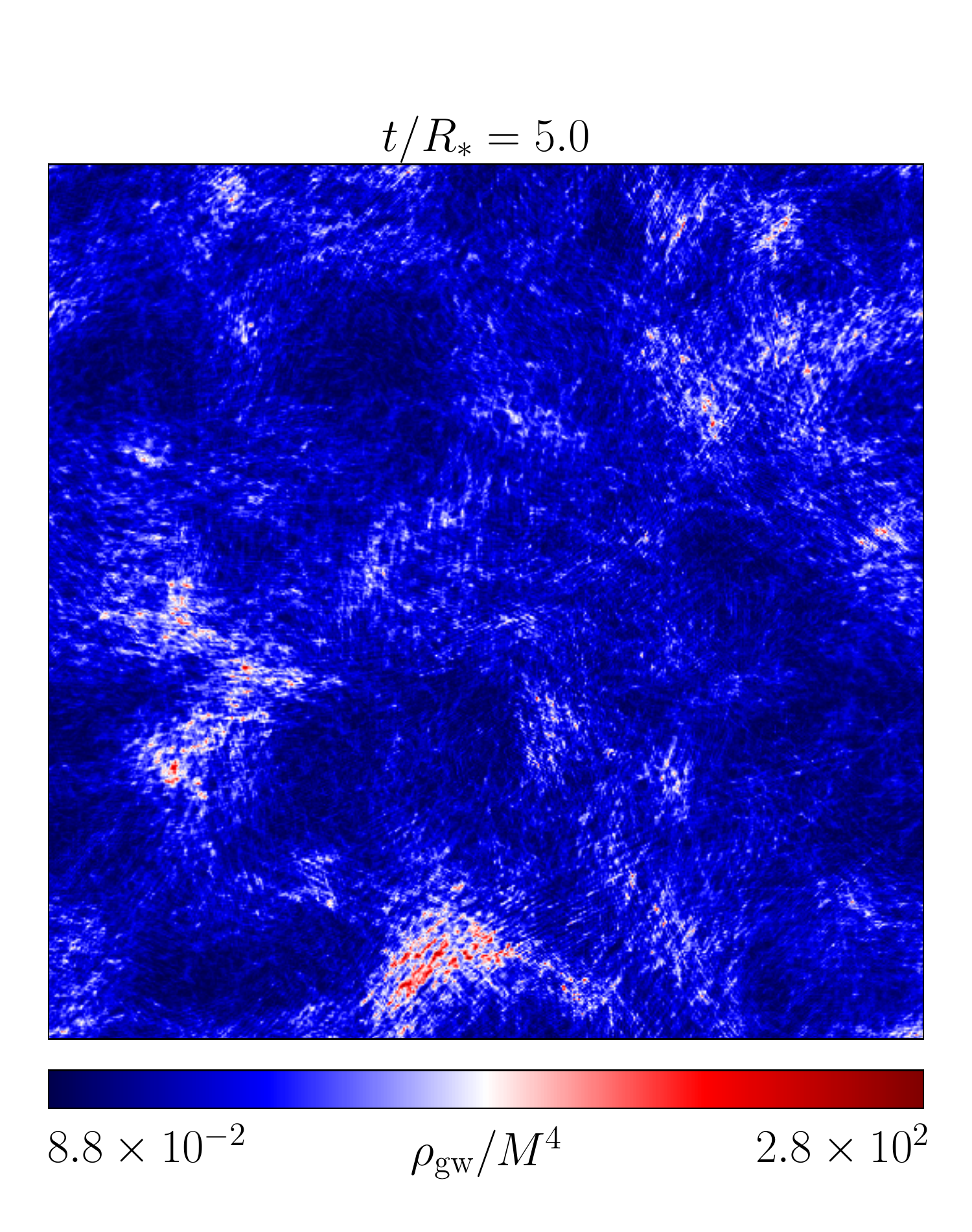}
	\end{tabular}
	\begin{tabular}{@{}c@{}}\hspace{-5mm}
		\includegraphics[width=.35\linewidth,trim={0 0 0 15mm},clip]{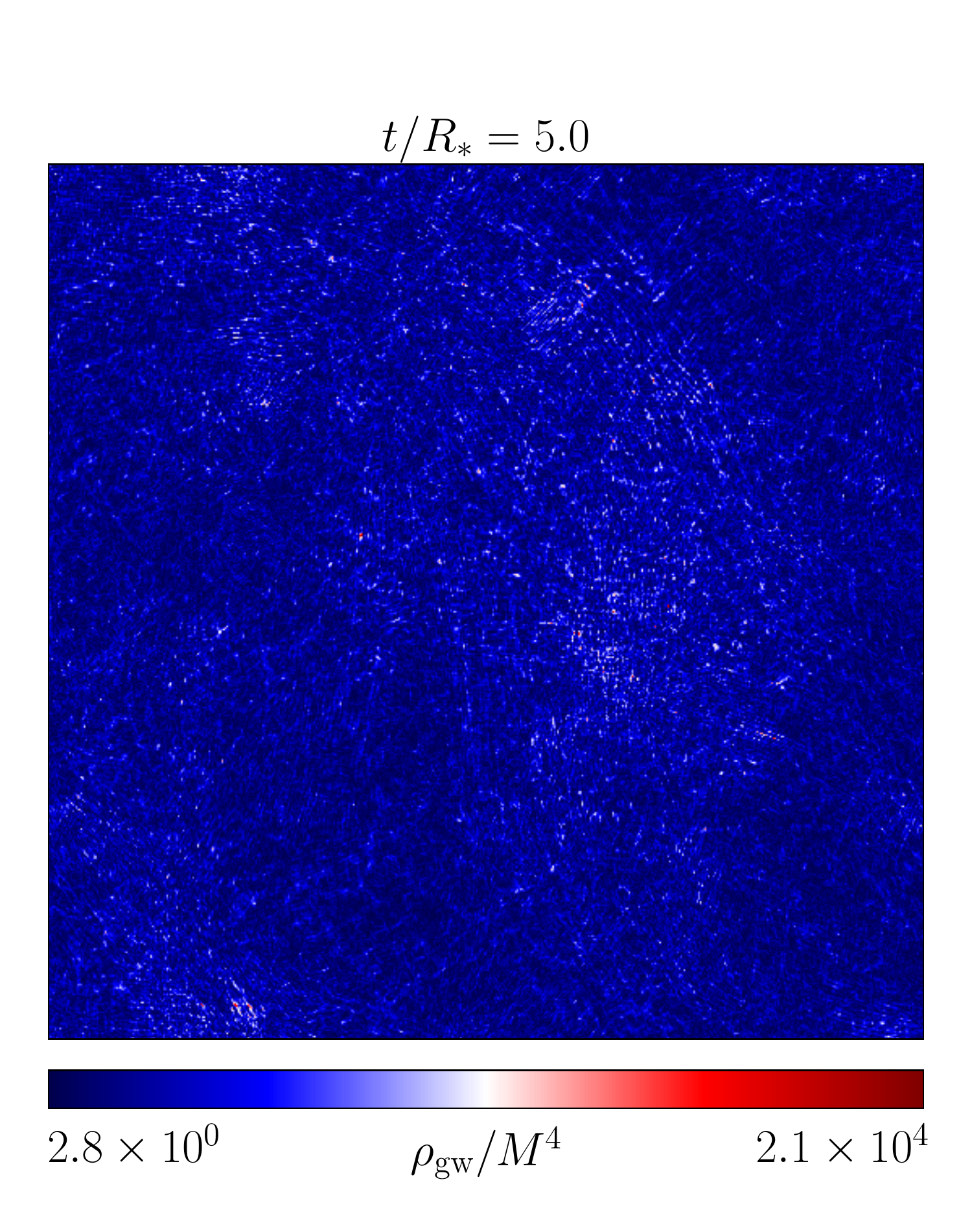}
	\end{tabular}
	
	\vspace{-2mm}
	
	\begin{tabular}{@{}c@{}}\hspace{-5mm}
		\includegraphics[width=.35\linewidth,trim={0 0 0 15mm},clip]{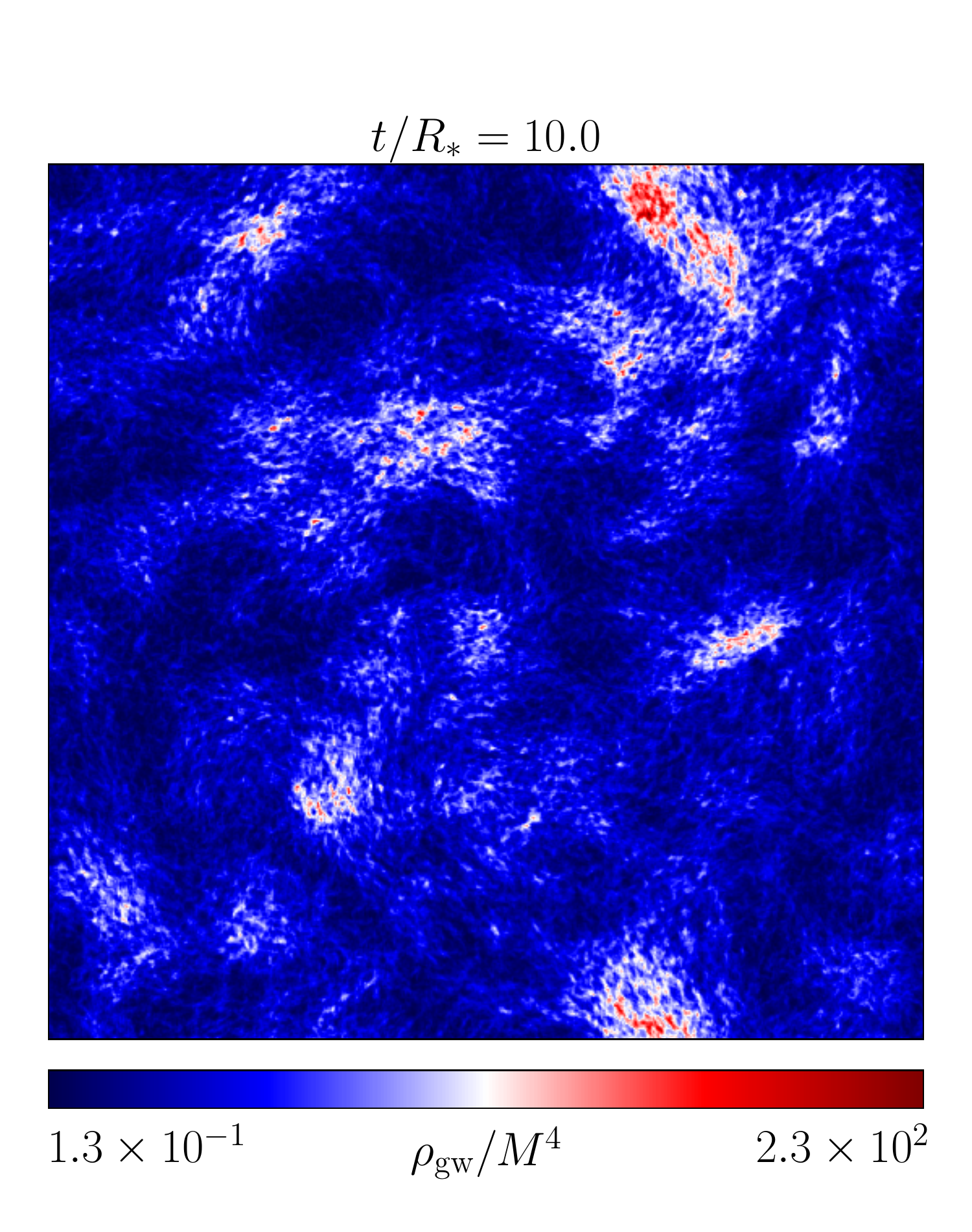}
	\end{tabular}
	\begin{tabular}{@{}c@{}}\hspace{-5mm}
		\includegraphics[width=.35\linewidth,trim={0 0 0 15mm},clip]{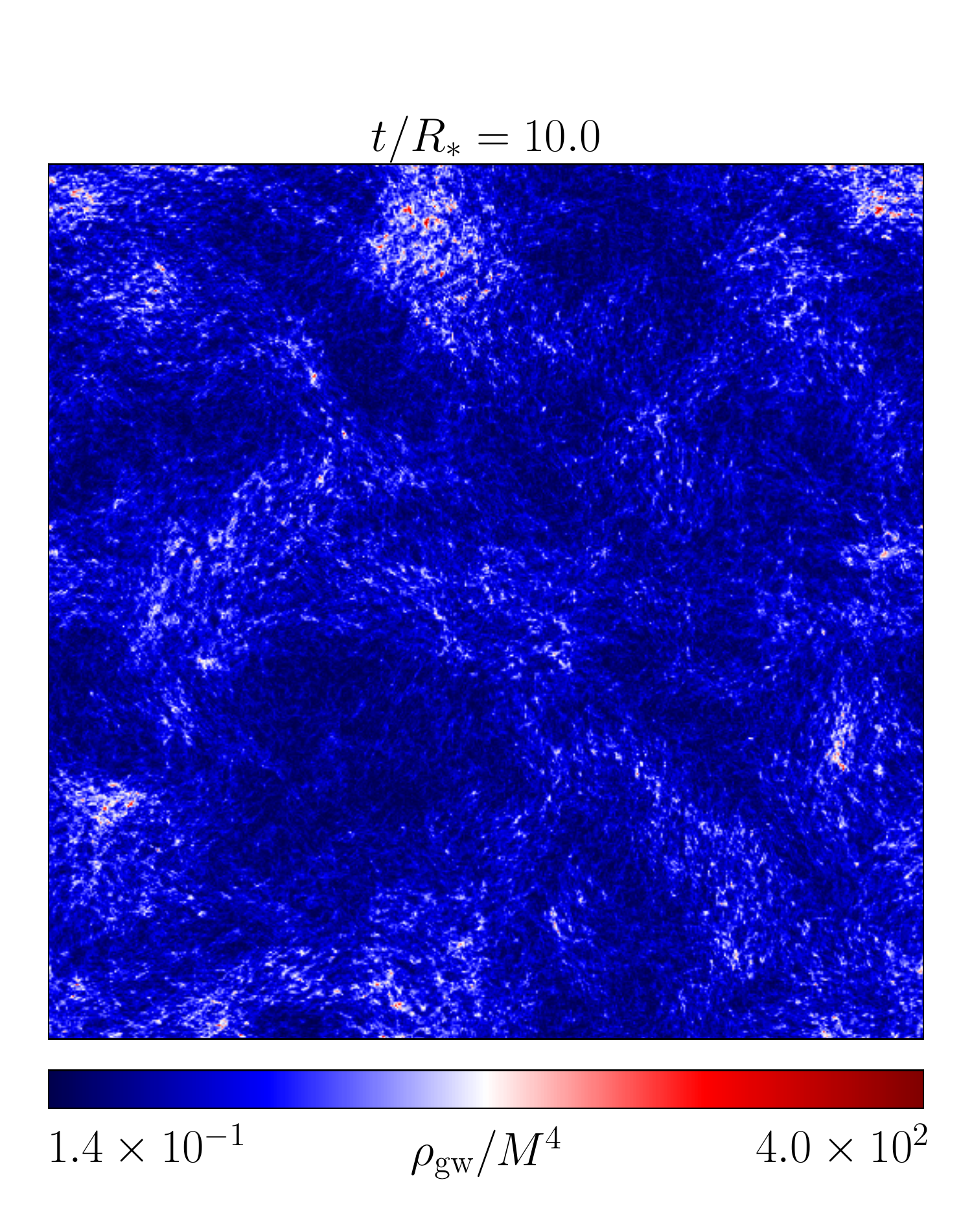}
	\end{tabular}
	\begin{tabular}{@{}c@{}}\hspace{-5mm}
		\includegraphics[width=.35\linewidth,trim={0 0 0 15mm},clip]{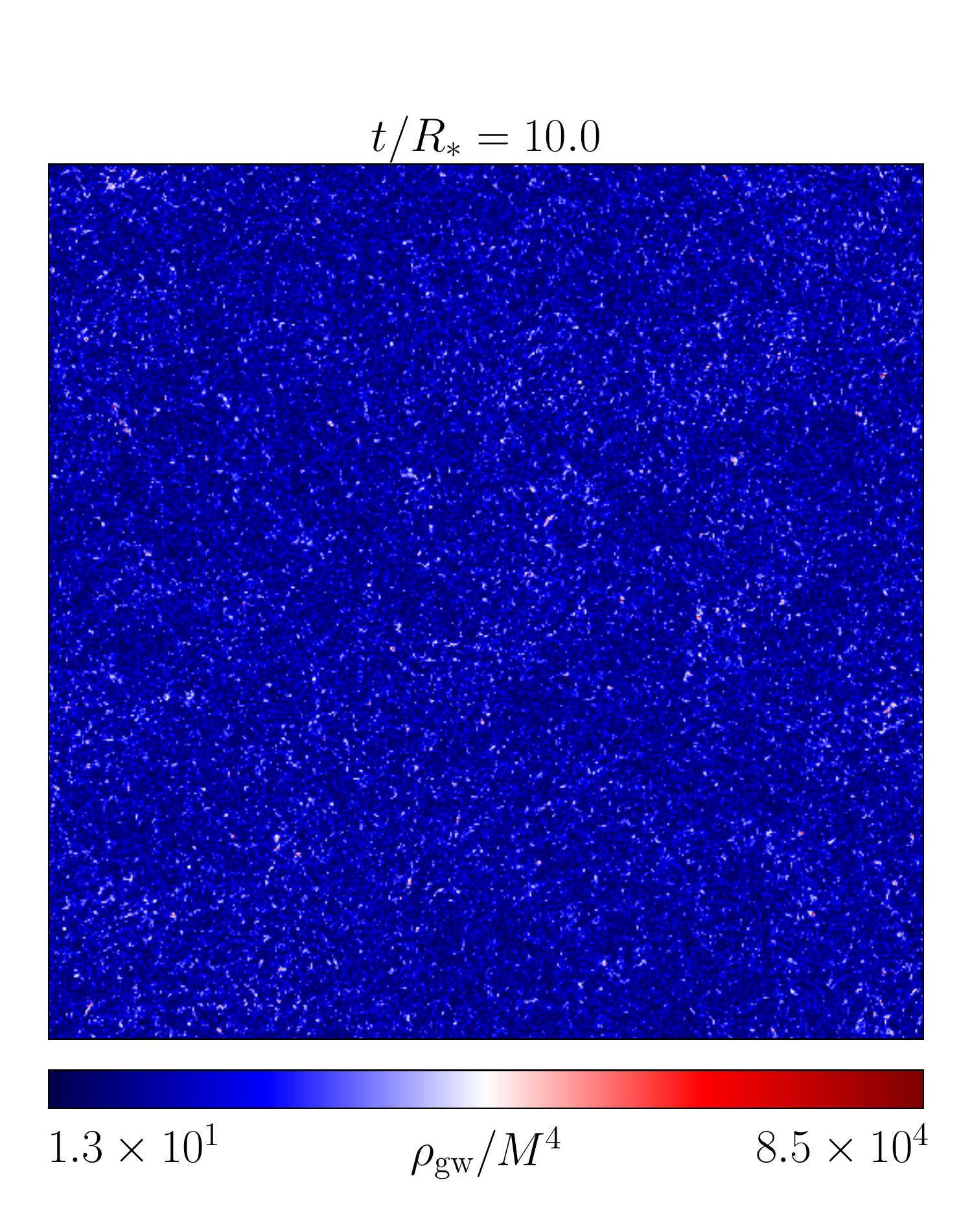}
	\end{tabular}
	
	\vspace{-2mm}
	
	\caption{Two dimensional slices of simulations described in the caption of Fig.\ \autoref{fig:pspec_gw} for $\xi=0$ (LEFT), $\xi=1 \times 10^{-4}$ (MIDDLE), and $\xi=1 \times 10^{-3}$ (RIGHT). The scales are specific to all plots, therefore, the color codes differ in each graph. Both axes have length $L=N dx$ with $N=1024$ and $dx=0.22$. Numbers in the color bars indicate maximum and minimum values of GW energy density in that particular slice.}
\label{fig:nmc_2D_snapshots}
\vspace{0mm}
\end{figure*}

We have run three main simulations for different $\xi$ values and for $\xi=0$, results of which are shown in Fig.\ \autoref{fig:pspec_gw} in terms of GW power spectra. Additionally, two dimensional slices of the simulation boxes are given in Fig.\ \autoref{fig:nmc_2D_snapshots} for three different runs showing only the spatial distribution of GW energy density for a clearer representation. The values of the coupling constant have been chosen such that the deviation is small in comparison and they are compatible with the constraints found in Refs.\ \cite{Bezrukov2007,Hrycyna2015,Arapoglu2019,Akin2020,Figueroa2021} from different contexts. We have used only positive values of $\xi$ since negative values of the coupling constant are forbidden within the interested interval as indicated in Sec.\ \autoref{sec:effective_potential}. Outside that region, on the other hand, negative values give the same potential shape which can already be achieved with some positive value, therefore, the outcome would be no different.

The results for $\xi=0$, which are equivalent to the ones originally found in Ref.\ \cite{Cutting2018}, have been produced to check that our code yields compatible outcomes and to complete our analysis in this paper. In addition to that we also provide the results of two different simulations for $\xi=0$ in the appendix for a consistency check of our code. As mentioned in the previous section we have numerically solved Eq.\ \autoref{eq:eom_phi_numerical} and Eq.\ \autoref{eq:eom_uij_numerical} with the leapfrog algorithm in all simulations for the time advance. We have implemented numerical values for the parameters in the first line of Table \autoref{tab:bubble_profile}. The simulations have been performed with simultaneous nucleation that is all the bubbles have been nucleated in the beginning of each run. Within the scope of the current study we have seen no necessities to conduct simulations with different nucleation types since we intended to compare only the immediate effect of the non-minimal coupling which should be similar for the other kinds as well.

As seen from Fig.\ \autoref{fig:pspec_gw} (and from the ones represented in the appendix), regarding to the case of $\xi=0$, the results coincide well with the outcomes of Refs.\ \cite{Cutting2018,Cutting2020}. On the other hand, for the simulations with $\xi \neq 0$ the first thing to notice from the power spectrum graphs is that the highest peak moves from a frequency associated with the mean bubble separation, $R_*$, towards the one closed to the bubble wall width, $l_0$, which is also related to the scalar field mass in the broken phase. If we increase the value of the coupling constant, this difference becomes apparent more quickly. In addition to growth of the peak connected with $l_0$, the peak around $R_*$ also rises in course of time with an increasing $\xi$ while its location is almost the same in all simulations around $t/R_*=1$. 

We have given two dimensional snapshots in Fig.\ \autoref{fig:nmc_2D_snapshots} for three simulations with $\xi=0$, $\xi=1 \times 10^{-4}$, and $\xi=1 \times 10^{-3}$. The effect of the friction term in Eq.\ \autoref{eq:eom_uij_numerical} creates a discernible pattern in the figures since that term suppresses the oscillations more with a higher value of the coupling constant which also causes the GW energy density to increase seemingly. While it is possible to see the traces of $R_*$ in figures for $\xi=0$, and $\xi=1 \times 10^{-4}$ at relatively late times, the same thing does not apply to their counterpart for $\xi=1 \times 10^{-3}$ where $l_0$ prevails over $R_*$ quickly as seen from its power spectra. Furthermore, the spatial distribution becomes very homogeneous in comparison with a significant increase in amplitudes that was also reported in Refs.\ \cite{Oikonomou2022,Odintsov2022} for a slightly different context.

\begin{figure*}[!t]
    \centering

    \begin{tabular}{@{}c@{}}
        \includegraphics[width=.5\linewidth]{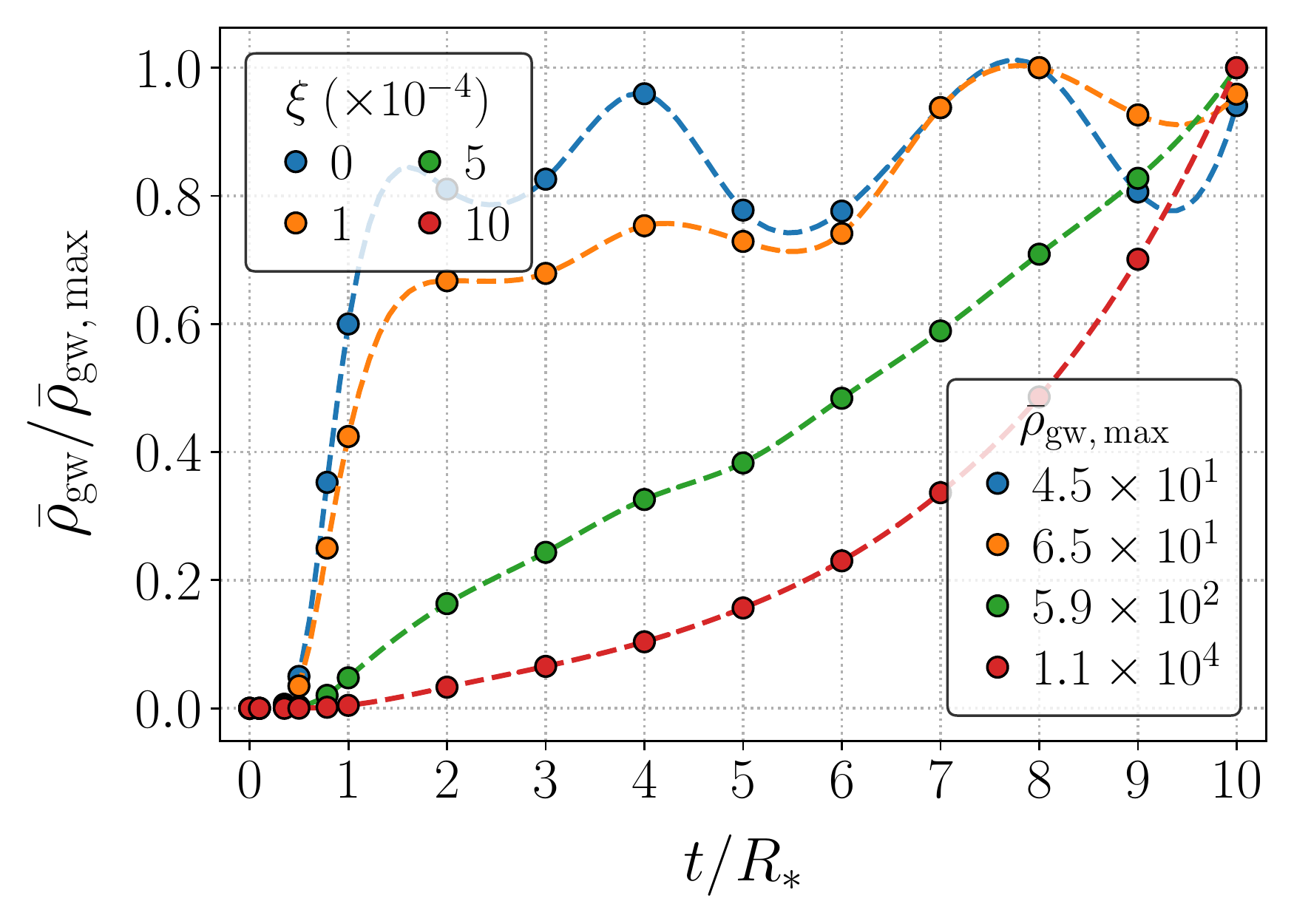}
    \end{tabular}

    \caption{Time evolution of the average GW energy densities normalized by individual maximum values to be able to show them in the same graph. The data points have been extracted from the simulations whereas the dashed lines are simple fits to guide the eye.}
\label{fig:densities} \vspace{3mm}
\end{figure*}

Regarding to time evolution of the GW energy densities defined in Eq.\ \autoref{eq:gw_density}, as seen from Fig.\ \autoref{fig:densities} the oscillations are obvious after the bubble collision phase for $\xi=0$ and $\xi=1 \times 10^{-4}$ with a slightly increasing trend. However, the other two data sets show that they continue to increase exponentially in the same time scale. Based on the results in power spectra it could be said that when the peak related to $l_0$ is higher than the one for $R_*$, oscillatory behavior in GW energy density does not appear in current outcomes. This means that exponentially increasing part of the solution becomes more dominant in comparison with the component controlling the oscillations. Nevertheless, together with the inclusion of modification in equation of motion for the scalar field, the expansion of the Universe should alter the situation since the Hubble parameter comes into play in the coefficient of the friction term in Eq.\ \autoref{eq:eom_uij_numerical}. Moreover, the reason for the increment in amplitudes as mentioned in the previous paragraph is evident now from the time evolution of GW energy densities.

\newpage

\section{CONCLUSION\label{sec:conclusion}}

In this work we have examined the effect of non-minimally coupled scalar field on the power spectrum of GW energy density in first-order vacuum phase transitions. To this end, we have succinctly analyzed the equations in an expanding background which led to the definition of an effective potential through Eq.\ \autoref{eq:effective_potential}. After integrating that expression, we have found the form of the effective potential which determines the behavior of the scalar field. Then, we have performed three dimensional numerical simulations in order to calculate GW power spectra. 

We have also briefly reviewed the initial bubble profile and we have seen that it is still a good approach to apply the thin-wall approximation in the parameter space that we are interested in. We have presented the characteristic parameter values of our model in Table \autoref{tab:bubble_profile} for different coupling constants. It seems that if we increase the value of the coupling constant, the critical radius and the wall width also increase as their ratio shows the same behavior which means that the system approaches to the regime where the thin-wall approximation can be applied. On the other hand, both value of the scalar field in true vacuum and the vacuum energy density decrease responding to diminishing height of the potential barrier which occurs as $\xi$ increases. This situation continues up to a critical value of $\xi$ until two minima become degenerate and then it goes the other way around.

The coupling constant has a significant effect on the characteristic length scales of the model as mentioned above. However, the values of $\xi$ should be small in comparison with the ones given in Table \autoref{tab:bubble_profile} if we consider the limitations coming from different subjects as suggested by the previous studies \cite{Bezrukov2007,Hrycyna2015,Arapoglu2019,Akin2020,Figueroa2021}. Our work in this paper also partially justifies the results in the literature as seen from the figures of GW power spectra and time evolution of GW energy densities. Moreover, it can be possible to find a restriction on $\xi$ considering the expansion of the Universe in the model as well. 

The potential of the non-minimally coupled scalar field can be written effectively as $V_{\rm{eff}}(\phi) = V(\phi) + \xi \phi \R$. Since the Ricci scalar contains only the scale factor and its derivatives in FLRW universe, this term can be considered as if it makes a contribution to the mass of the scalar field, an effective mass in a manner of speaking. Except for the investigation of initial bubble profile, even though we did neglect this effect in the simulations, the change in the highest peak of GW power spectra reflects the imprint of $\xi$ on the phase transitions and signals the necessity of further studies including such effects that might also provide insights for the Higgs inflation scenarios.

It has been known that the gravitation and the vacuum decay have mutual non-negligible effects on each other \cite{Coleman1980} let alone the non-minimal coupling. Nevertheless, in this work we have intended to determine the impact level of the non-minimal coupling on the power spectra of GW energy density by covering a limited time interval. Our results indicate that this well-motivated model should be investigated in detail with more realistic setups in accordance with the conditions in the early Universe.

\clearpage

\clearpage

\section*{Acknowledgements}

This work is supported by the Scientific and Technological Research Council of Türkiye (TÜBİTAK) through grant number 121F066. Computing resources used in this work were provided by the National Center for High Performance Computing of Turkey (UHeM) under grant number 5013072022. The code used for the numerical calculations reported in this paper were tested through the simulations performed at TUBITAK ULAKBIM, High Performance and Grid Computing Center (TRUBA resources).

\bibliographystyle{apsrev4-2}
\bibliography{references}

\newpage

\section*{Appendix\vspace{-3mm}}

Here we provide the results of two different simulations to show that our code produces consistent outcomes with the previous ones obtained in Refs.\ \cite{Cutting2018,Cutting2020}. To do that we set $\xi=0$ as it corresponds to the case in the mentioned papers. In this work we have followed Ref.\ \cite{Cutting2018} in the context of the values for the simulation parameters. However, the case $\Bar{\lb}=0.84$ in Ref.\ \cite{Cutting2020} also coincides with the parameter values at the top row of Table \autoref{tab:bubble_profile} that have been used through the simulations results of which are given below.

\begin{figure*}[!h]
    \centering

    \begin{tabular}{@{}c@{}}\hspace{-7mm}
	\includegraphics[width=.5\linewidth]{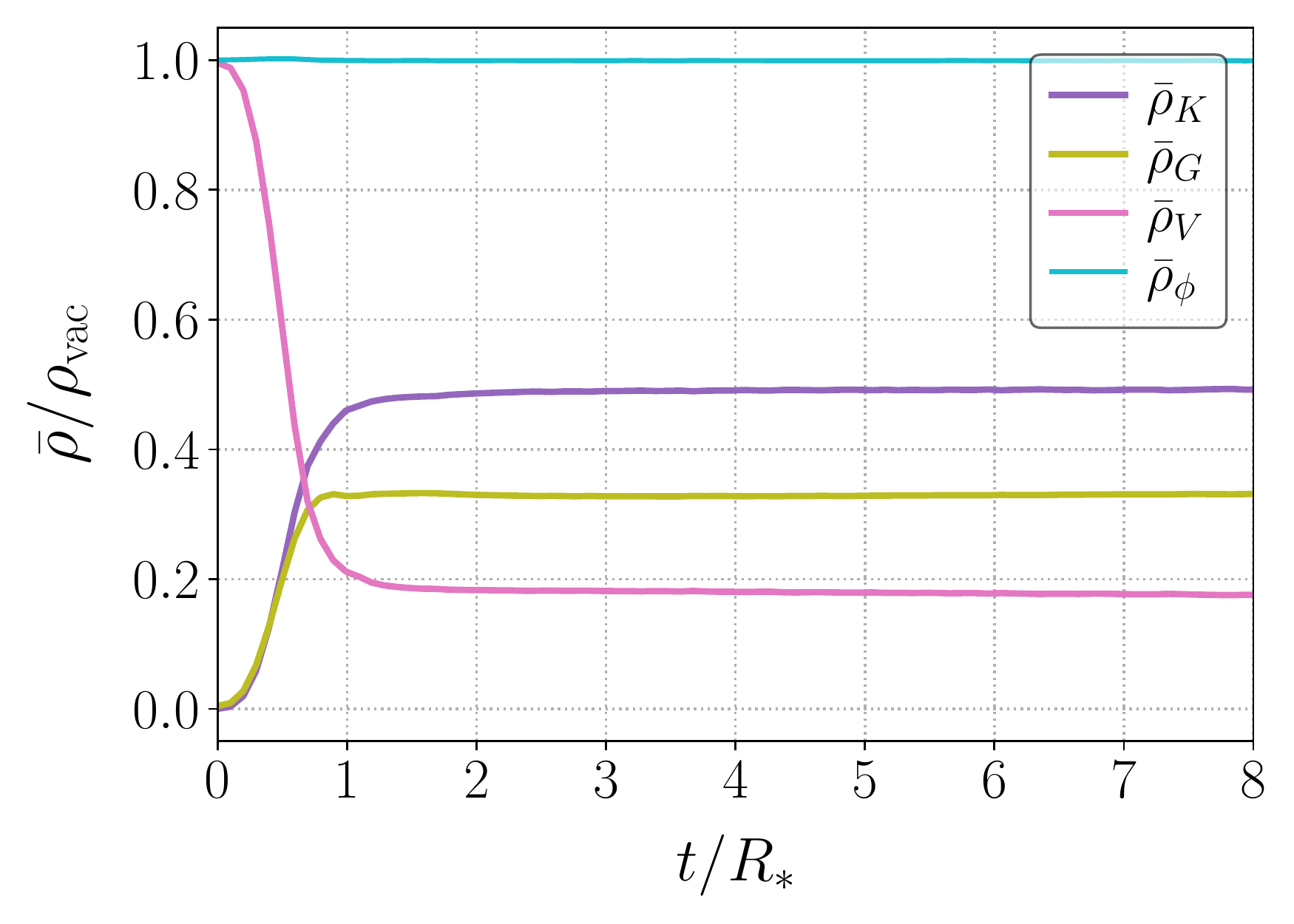}
    \end{tabular}

    \vspace{-3mm}

    \caption{Time evolution of the scalar field mean densities for $\xi=0$ as defined in Eq.\ \autoref{eq:scalar_densities}.}
\label{fig:scalardensities} \vspace{0mm}
\end{figure*}

To begin with we plot time evolution of the scalar field mean energy densities in Fig.\ \autoref{fig:scalardensities} where kinetic, gradient, and potential energy densities are defined as
\begin{equation}
    \rho_{\rm{K}} = \sfrac{1}{2} \dot{\phi}^2 \:, \qquad \rho_{\rm{G}} = \sfrac{1}{2} \big( \nb \phi \big)^2 \:, \qquad \rho_{\rm{V}} = V(\phi) - V(\phi_{\rm{t}}) \:.
\label{eq:scalar_densities}
\end{equation}
We see from the figure that the total energy of the scalar field initially forms from its potential as the kinetic and the gradient energies are close to zero. This situation alters until around $t/R_*=1$ after which they all settle down to nearly a constant value. Comparison of Fig.\ (4) in Ref.\ \cite{Cutting2018} and Fig.\ (6) in Ref.\ \cite{Cutting2020} with Fig.\ \autoref{fig:scalardensities} below shows that our code works properly in this manner.

\begin{figure*}[!h]
    \centering

    \begin{tabular}{@{}c@{}}\hspace{-5mm}
	\includegraphics[width=.48\linewidth]{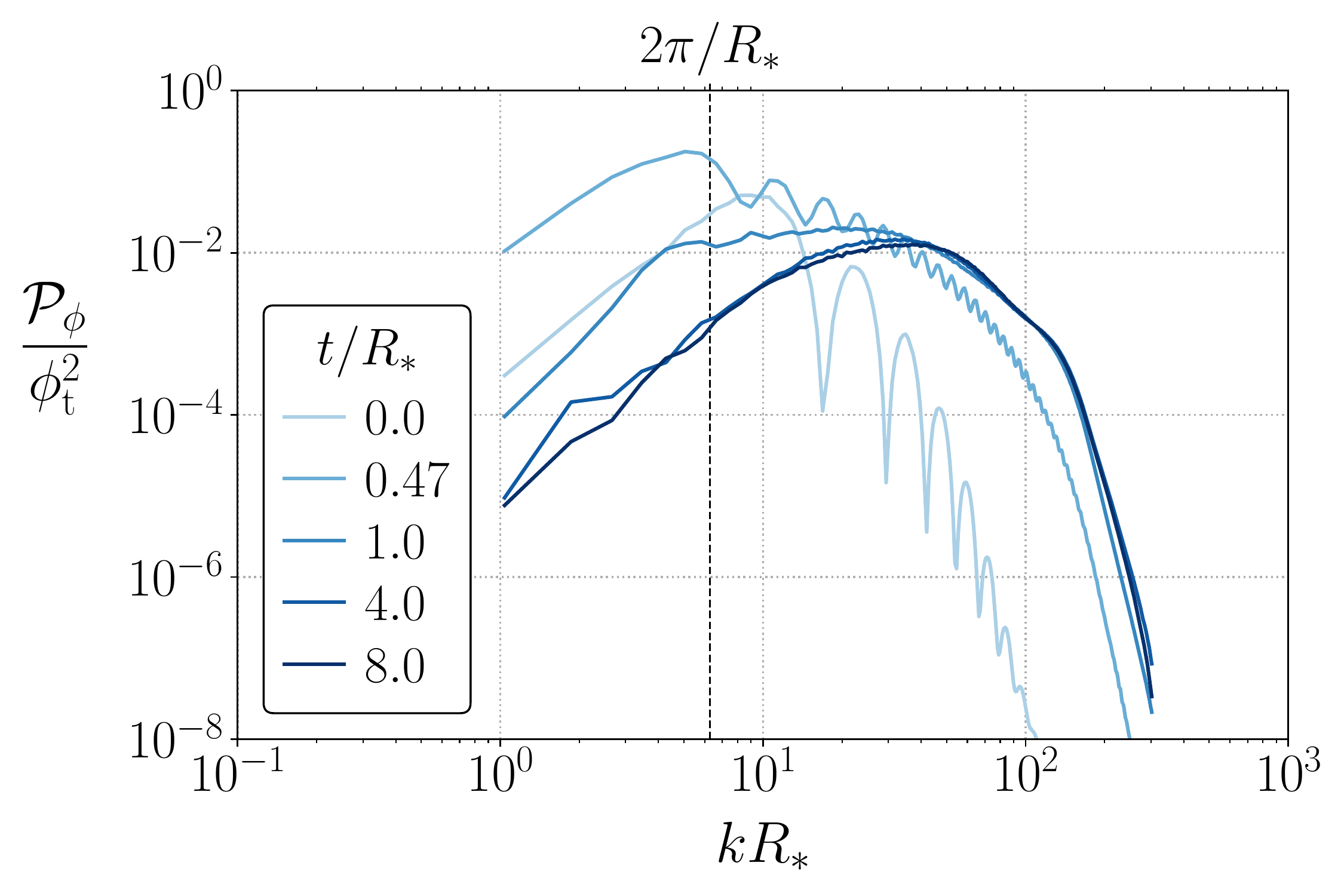}
    \end{tabular}
    \begin{tabular}{@{}c@{}}
	\includegraphics[width=.48\linewidth]{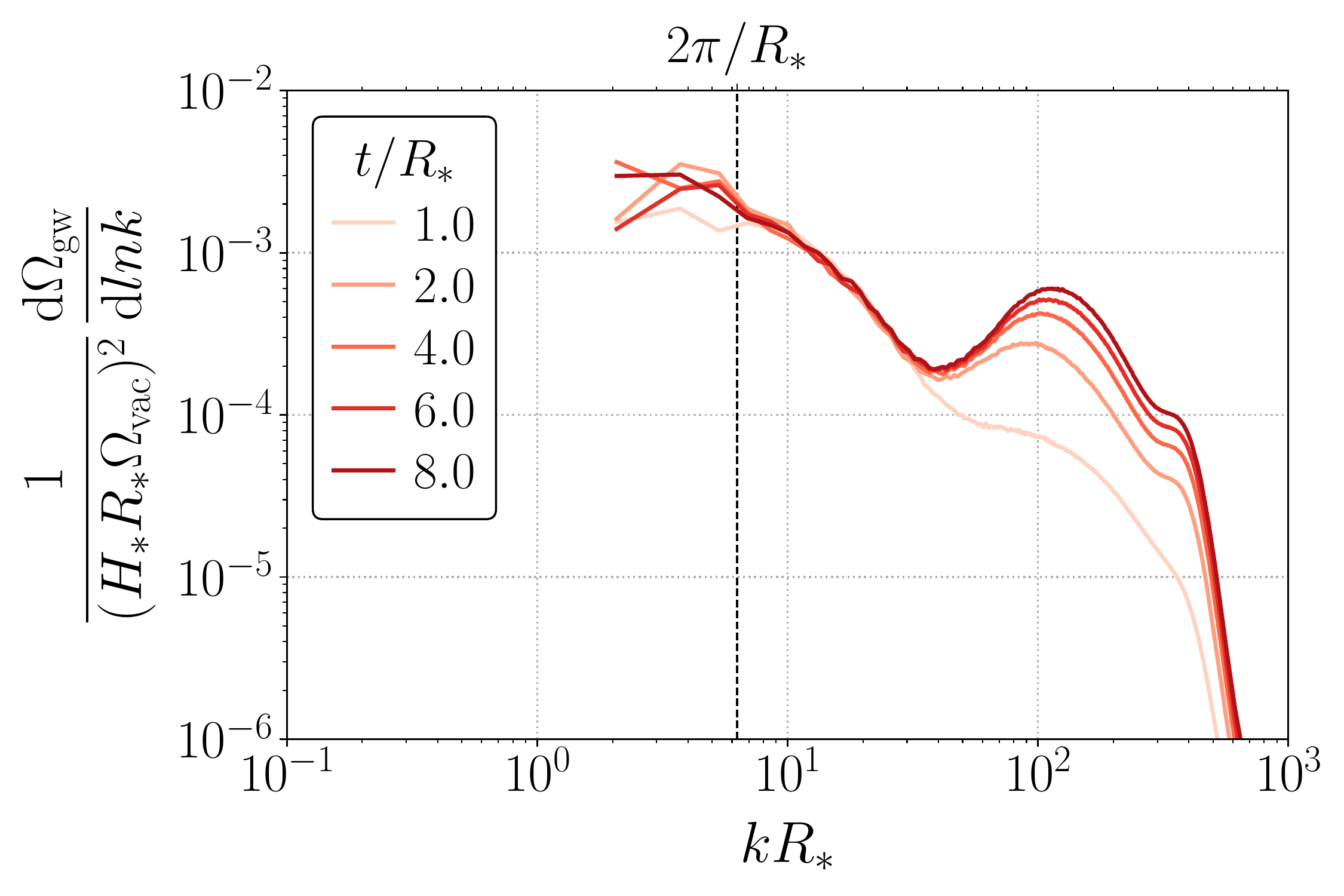}
    \end{tabular}

    \vspace{-2mm}

    \caption{(LEFT) The power spectrum of the scalar field and (RIGHT) the power spectrum of the GW energy density.}
\label{fig:pspecs_appendix} \vspace{0mm}
\end{figure*}

\begin{figure*}[!h]
    \centering

    \begin{tabular}{@{}c@{}}\hspace{-5mm}
	\includegraphics[width=.5\linewidth]{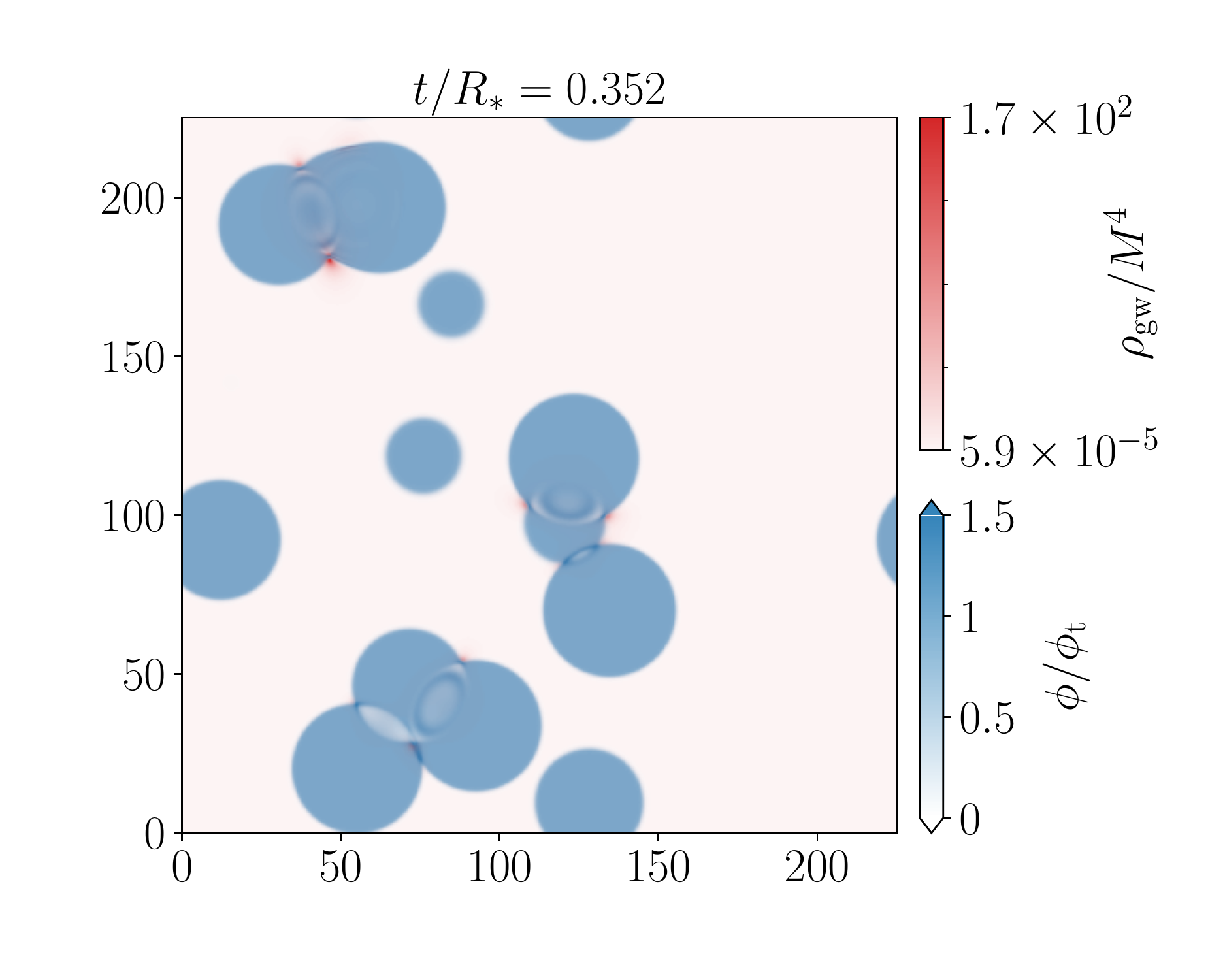}
    \end{tabular}
    \begin{tabular}{@{}c@{}}\hspace{-3mm}
	\includegraphics[width=.5\linewidth]{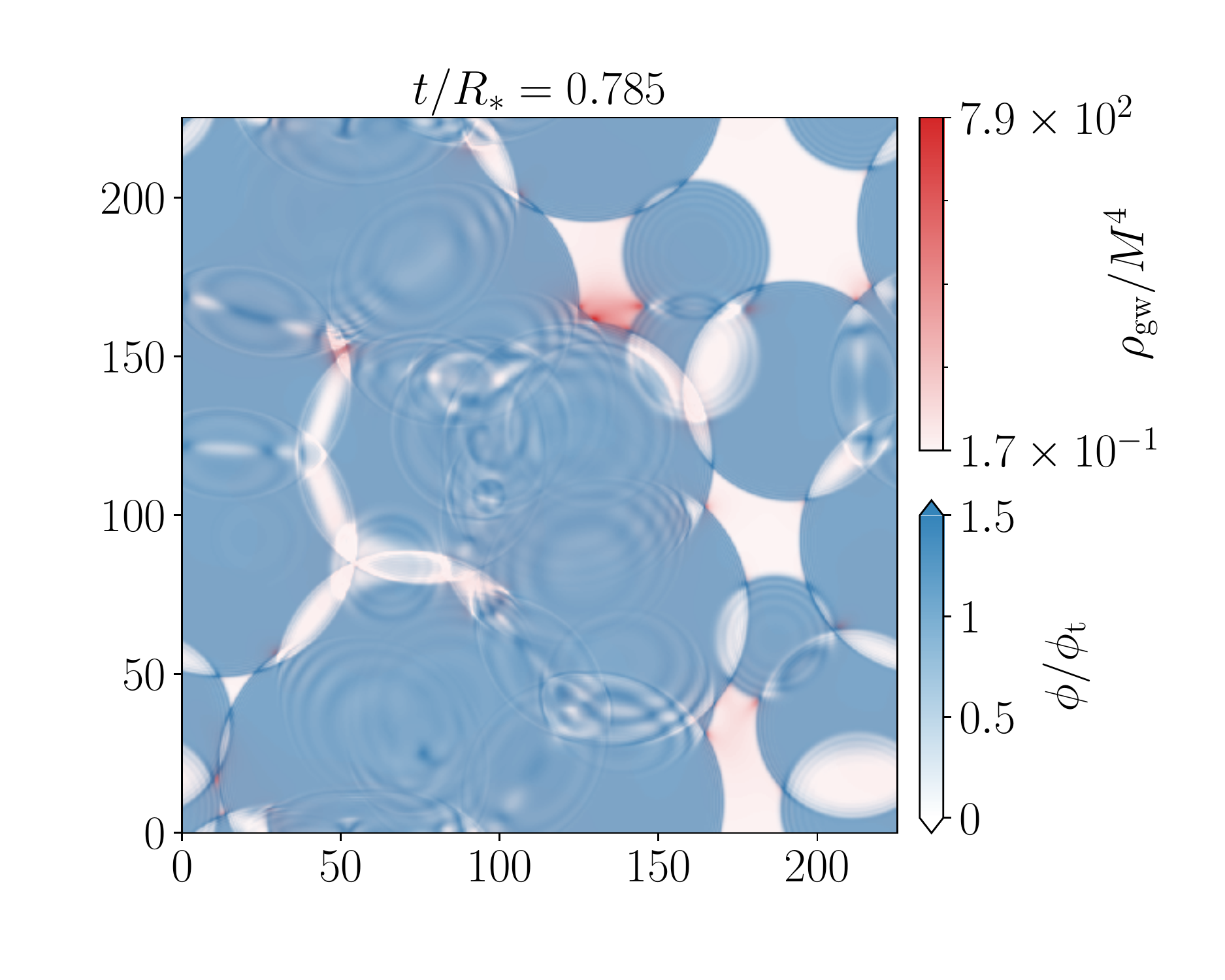}
    \end{tabular}

    \vspace{-5mm}

    \begin{tabular}{@{}c@{}}\hspace{-7mm}
	\includegraphics[width=.5\linewidth]{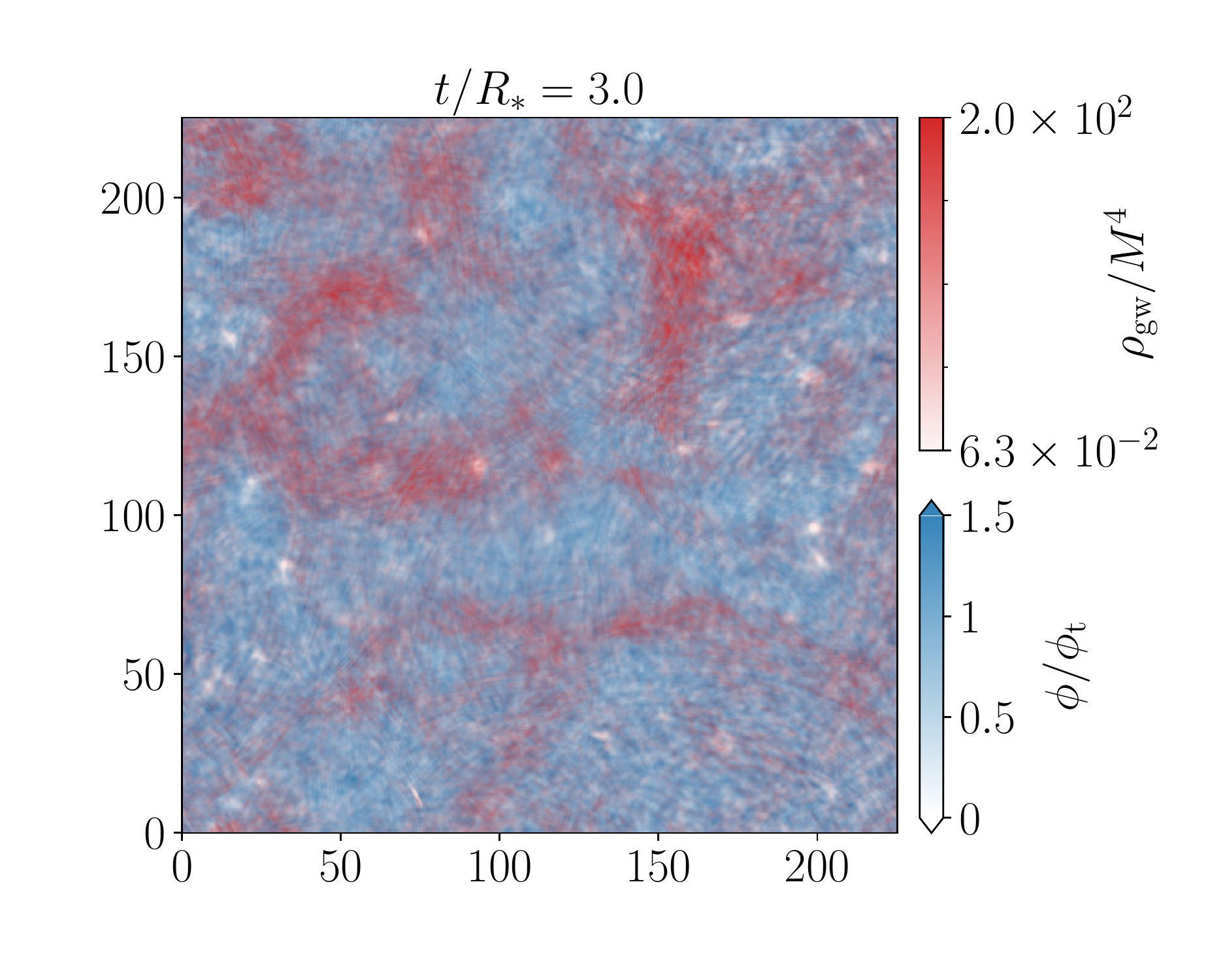}
    \end{tabular}
    \begin{tabular}{@{}c@{}}\hspace{-7mm}
	\includegraphics[width=.5\linewidth]{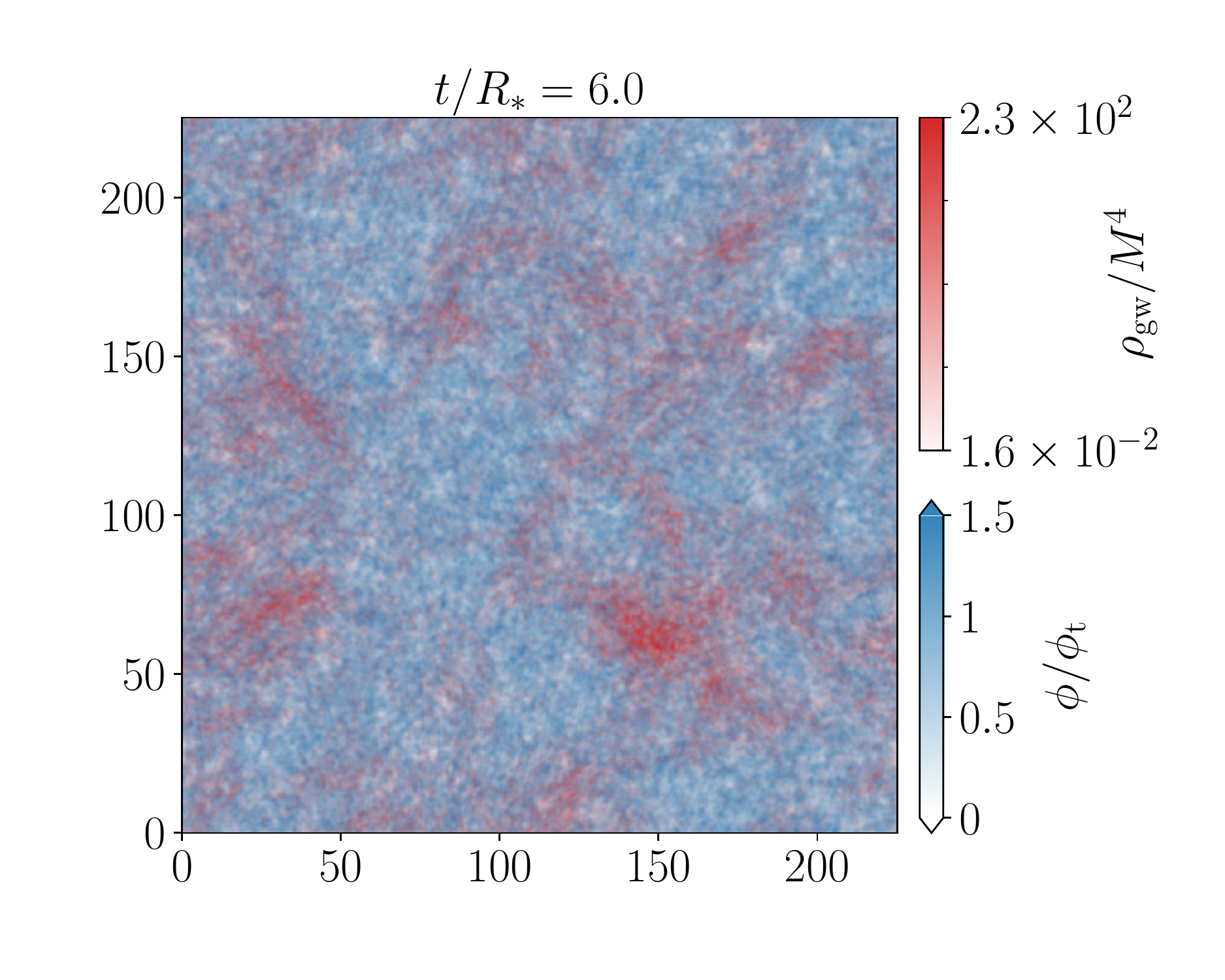}
    \end{tabular}

    \vspace{-3mm}

    \caption{Two-dimensional slices of a simulation with $N=1024$ and $dx=0.22$ for $\xi=0$.}
\label{fig:snaps_appendix} 
\end{figure*}

Another important result to check is the power spectrum which we present in Fig.\ \autoref{fig:pspecs_appendix} for both the scalar field and the GW energy density. The power spectrum of the scalar field and the power spectrum of the GW energy density correspond to Fig.\ (6a) and Fig.\ (8) in Ref.\ \cite{Cutting2018}, respectively. We have used the same parameters with the mentioned paper, that is, to produce the outcome of the power spectrum for the scalar field (the GW energy density) we have taken $N=512$ ($N=1024$), $dx=0.44$ ($dx=0.22$), and 512 (64) bubbles initiated simultaneously in the beginning of the simulation. Comparison of the results show that our code gives compatible outputs for the power spectrum as well. On the other hand, although it is not as important as the previous results in order to validate the consistency of our code directly, in Fig.\ \autoref{fig:snaps_appendix} we also present two-dimensional slices obtained from the same simulation that gave the power spectrum of the GW energy density. These snapshots can be compared with Fig.\ (2) in Ref.\ \cite{Cutting2018}.

\end{document}